\documentclass[11pt,a4paper]{article}
\pdfoutput=1

\usepackage{jheppub}

\usepackage{latexsym,amssymb,amsmath,amsfonts,amstext}
\usepackage{slashed}
\usepackage{mathrsfs}
\usepackage{bbold}
\usepackage{xcolor}
\usepackage[toc,page]{appendix}

\def\Slash#1{\rlap{\hbox{$\mskip 3 mu /$}}#1}      	
\def\oneone{\rlap 1\mkern4mu{\rm l}} 			

\title{On the quantum entropy function  in 4d gauged supergravity}

\author{Kiril Hristov$^a$,}
\author{Ivano Lodato$^b$,}
\author{and Valentin Reys$^{c,d}$}

\affiliation{$^a$ Institute for Nuclear Research and Nuclear Energy, \\ Bulgarian Academy of Sciences, 1784 Sofia, Bulgaria}
\affiliation{$^b$ Department of Physics and Center for Field Theory and Particle Physics, \\ Fudan University, 200433 Shanghai, China}
\affiliation{$^c$ Dipartimento di Fisica, Universit\`a di Milano--Bicocca, \\ Piazza della Scienza 3, I-20126 Milano, Italy}
\affiliation{$^d$ INFN, sezione di Milano--Bicocca, \\ Piazza della Scienza 3, I-20126 Milano, Italy}

\emailAdd{khristov@inrne.bas.bg, ilodato@fudan.edu.cn, valentin.reys@unimib.it}

\abstract{
We analyze BPS black hole attractors in the conformal 4d gauged supergravity formalism and apply the technique known as supergravity localization in order to evaluate Sen's quantum entropy function \cite{Sen:2008vm} in the AdS$_2 \times$S$^2$ near-horizon geometry. Under certain assumptions, we reduce the exact expression of the functional integral to a finite-dimensional integral for a number of supersymmetric black holes in gauged supergravity with AdS asymptotics subject to a holographic description via a dual field theory. Examples include the asymptotically AdS$_4 \times$S$^7$ Cacciatori-Klemm black holes \cite{Cacciatori:2009iz} in M-theory and the asymptotically AdS$_5 \times$S$^5$ generalizations of Gutowski-Reall black holes \cite{Gutowski:2004ez} and Benini-Bobev black strings \cite{Benini:2013cda} in type IIB, as well as the recently constructed asymptotically AdS$_4 \times$S$^6$ solutions \cite{Guarino:2017eag,Guarino:2017pkw} in massive type IIA. Our results provide an important first step towards a gravitational counterpart to the exact evaluation of supersymmetric partition functions at finite $N$ for the holographically dual field theories in these examples.
}

\begin{document}

\maketitle

\section{Introduction, main procedure and summary of results}
\label{sec:intro}

The understanding of the microscopic origin of black hole entropy is one of the important problems that only a consistent theory of quantum gravity can answer. Historically this problem has led to some major developments in string theory, starting with the work of Strominger and Vafa \cite{Strominger:1996sh} where they computed the leading Bekenstein-Hawking entropy of a class of asymptotically flat black holes. This breakthrough served as one of the inspirations for the discovery of the AdS/CFT correspondence by Maldacena \cite{Maldacena:1997re}, and spurred an extensive research activity culminating in exact microscopic state counting formulae in string theories compactified on tori and K3 surfaces \cite{Maldacena:1999bp,Dijkgraaf:1996xw}. It also shed light on microstate counting in string theories compactified on general Calabi-Yau 3-folds \cite{Denef:2007vg}. Macroscopically, the exact evaluation of the entropy of certain asymptotically flat black holes in maximal supergravity, including all possible quantum corrections, was conducted in \cite{Dabholkar:2010uh,Dabholkar:2011ec,Dabholkar:2014ema} and shown to agree with the corresponding microscopic counting. The success of this enterprise relied on several powerful techniques developed in well-known works such as \cite{Pestun:2007rz,Sen:2008vm}. The literature on this topic is extensive and we refer the reader to the work \cite{Dabholkar:2010uh} for a list of relevant references.  

A more recent development in the field of microscopic entropy counting was initiated in \cite{Benini:2015eyy} where the Bekenstein-Hawking entropy of supersymmetric asymptotically AdS$_4$ black holes \cite{Cacciatori:2009iz} was reproduced by a large $N$ evaluation of the topologically twisted index \cite{Benini:2015noa} in the holographically dual ABJM theory \cite{Aharony:2008ug}. Many related developments and generalizations, some of which will be relevant to this work, followed both on the gravity and on the field theory side \cite{Hosseini:2016tor,Hosseini:2016ume,Benini:2016hjo,Hosseini:2016cyf,Cabo-Bizet:2017jsl,Hosseini:2017mds,Liu:2017vll,Azzurli:2017kxo,Jeon:2017aif,Hosseini:2017fjo,Benini:2017oxt,Bobev:2017uzs,Halmagyi:2017hmw,Liu:2017vbl,Cabo-Bizet:2017xdr,Toldo:2017qsh,Bobev:2018uxk}. Perhaps a natural endpoint of this research topic is again the exact evaluation of the quantum entropy for this class of supersymmetric black holes. In some of the references cited above, one can already find explicit results for the first quantum corrections both on the field theory and the gravity side. To evaluate the full exact quantum entropy however, one should consider separately the two sides of the duality since the challenges that one faces along the way are fundamentally different.

The natural quantity to consider on the field theory side is the topologically twisted index \cite{Benini:2015noa}, which is in principle a well-defined quantum object for any value of the coupling constant or the rank of the gauge group. By definition, this index counts supersymmetric states with a certain weight, and via the statistical arguments given in \cite{Benini:2016rke} it corresponds in the large $N$ limit to the saddle-point approximation for the degeneracy of states. This means that the macroscopic Bekenstein-Hawking entropy corresponds to an extremization of the index in a way that can be made very precise \cite{Benini:2015eyy,Benini:2016rke}. The full evaluation of the index at finite $N$, although well-defined, is unfortunately a technically very challenging problem.

The question of evaluating the exact quantum entropy is less clearly defined on the gravity side due to our lack of a fundamental understanding of gravity's quantum degrees of freedom. The leading Bekenstein-Hawking entropy is unquestionably related to states in the black hole interior, while the analysis of possible subleading contributions to the area law due to hair degrees of freedom living outside the horizon is already non-trivial (see \cite{Liu:2017vll,Liu:2017vbl}). In this paper we try to answer a slightly simpler question, namely how to evaluate at a full quantum level the contribution from gravitational states strictly inside the black hole by only considering the near-horizon geometry\footnote{We work with the general class of BPS near-horizon solutions relevant for the Cacciatori-Klemm asymptotically AdS$_4\times$S$^7$ black holes \cite{Cacciatori:2009iz} which automatically gives us access to a number of other interesting black hole embeddings in string theory. The general attractor mechanism in 4d gauged supergravity was analyzed in \cite{Cacciatori:2009iz,Dall'Agata:2010gj,Hristov:2010ri}, and later shown to be exhibited by: the BPS black strings with AdS$_5 \times$S$^5$ asymptotics in type IIB \cite{Benini:2013cda}, via a dimensional reduction to 4d \cite{Hristov:2014eza}; the BPS rotating black holes of Gutowski-Reall \cite{Gutowski:2004ez} and their generalizations \cite{Gutowski:2004yv,Chong:2005da,Chong:2005hr,Kunduri:2006ek} with AdS$_5\times$S$^5$ asymptotics, again via dimensional reduction to 4d \cite{Hosseini:2017mds}; and the recently found BPS attractors of massive type IIA theory compactified on an S$^6$ topology \cite{Guarino:2017eag,Guarino:2017pkw,Hosseini:2017fjo,Benini:2017oxt}. After analysing the quantum entropy function for the general class of near-horizon geometry, we will comment on each of these specific cases in section \ref{sec:examples}.\label{footnote1}}. 

We use the quantum entropy function formalism of Sen \cite{Sen:2008vm}, together with the supergravity localization technique used by Dabholkar, Gomes and Murthy in the case of asymptotically flat black holes \cite{Dabholkar:2010uh,Dabholkar:2011ec}. We will actually see that the technical tools needed on the gravity side have, to a large extent, already been developed in the off-shell supergravity literature \cite{deWit:1980lyi,deWit:1984wbb,deWit:1984rvr,deWit:2011gk}, but it is probably fair to say that a complete fundamental understanding of the localization procedure in gravity is still lacking. Therefore we first turn our attention towards carefully describing the logic behind our calculations, before summarizing the main results and commenting on their connection with field theory calculations.

\subsection{Localization procedure}
\label{subsec:sugra-localization}

We consider Sen's Quantum Entropy Function (QEF) \cite{Sen:2008vm}, defined as
\begin{equation}
\label{eq:macroscopic-degeneracy}
d_{\rm macro} (p^I, q_I) := \left< \exp \left( 4\pi\,q_I \int_0^{2\pi} W^I_{\tau}\,{\rm d} \tau \right) \right>_{\mathrm{AdS}_2}^{\mathrm{finite}}\ ,
\end{equation}
corresponding to the full Euclidean path integral of the gravitational theory on the AdS$_2$ space arising in the black hole near-horizon geometry, with an insertion of a Wilson line at the boundary. Here, $\tau$ is the periodic ``time'' coordinate of Euclidean AdS$_2$ (hereafter denoted as\ ${\cal H}_2$), the superscript `finite' denotes a regularization of the divergences arising from the non-compact nature of ${\cal H}_2$, and the $W^I_{\mu}$ are the real Euclidean $U(1)$ gauge fields under which the black hole is charged. The insertion of the Wilson line is needed so that we compute the macroscopic degeneracy in the microcanonical ensemble with fixed electric and magnetic charges $q_I$ and $p^I$ (in Planckian units), respectively. Note that the above formula makes sense at a full quantum level even if we lose the strict geometric meaning of the black hole space-times in the strong coupling regime of string theory. Thanks to the attractor mechanism \cite{Cacciatori:2009iz,Dall'Agata:2010gj,Hristov:2010ri}, $d_{\rm macro}$ is a function of the asymptotic charges only. These remain well-defined in the quantum regime, which is why it makes sense to use the microcanonical ensemble when evaluating the exact quantum degeneracy. 

Following the idea of \cite{Dabholkar:2010uh} and subsequent literature, here we want to use localization techniques in the bulk theory in order to simplify substantially the functional integral \eqref{eq:macroscopic-degeneracy} for the AdS$_2 \times$S$^2$ near-horizon geometries\footnote{Note that in general AdS black holes can have any Riemann surface $\Sigma_g$ for the topology of the horizon, see e.g.\ \cite{Caldarelli:1998hg}. Keeping the discussion general is fairly straightforward, and in the off-shell formalism this is explained in \cite{Hristov:2016vbm}. This does not lead to any complications, and therefore we just specialize to the spherical case in this paper for clarity of some formulas.} of supersymmetric black holes in four-dimensional gauged supergravity. Before describing the procedure step by step, let us first stress that we work in the conformal off-shell formalism of ${\cal N}=2$ Euclidean supergravity recently developed in \cite{deWit:2017cle}. This formalism has a clear advantage, because most of our analysis is done at the level of BPS variations and can therefore be performed without relying on the explicit form of the Lagrangian of the theory, allowing e.g. for the introduction of arbitrary higher-derivative terms that might arise from string theory corrections. 

Our main technical assumption is that we ``freeze'' the supermultiplet containing the graviton field, known as the Weyl multiplet. This means that our localizing Lagrangian, usually denoted as $Q_{\rm loc}V$, and the resulting critical locus $Q_{\rm loc} V|_{\mathrm{locus}} = 0$, does not involve fields from the Weyl multiplet. Therefore, up to issues in the one-loop determinant of $Q_{\rm loc}V$ and the integration measure along the locus which will be discussed in due course, our procedure becomes formally equivalent to the usual rigid (equivariant) supersymmetry localization on curved backgrounds where the matter content is provided by the supergravity theory.\\ 

Let us outline a general localization program for the problem we consider. In this paper we will explicitly perform all but the last of the following steps:

\begin{itemize}
\item {\it Step 1}: Write down the supersymmetric on-shell background solution in the supergravity theory we consider, repeating the analysis of \cite{deWit:2011gk,Hristov:2016vbm} in the Euclidean case. We fix both the supergravity multiplet and all matter multiplets to their BPS values which additionally solve the equations of motion\footnote{Compared to {\it rigid} supersymmetric localization this step is equivalent to putting the field theory on a curved space, but here we further impose the equations of motion since we are interested in evaluating an observable in a true gravitational background.}.

\item {\it Step 2}: Analyze the superalgebra of the fixed on-shell background, i.e.\ the attractor near-horizon geometry of BPS black holes in gauged supergravity, and pick a suitable localizing supercharge $Q_{\rm loc}$ which squares to a compact isometry of ${\cal H}_2$. 

\item {\it Step 3}: Determine the bosonic localization locus resulting from the chosen supercharge and associated localizing Lagrangian. For the usual choice of fermionic functional $V = \sum_\alpha \bar{\psi}_\alpha\,Q_{\rm loc}\psi_\alpha$, where the sum runs over all fermion fields of the theory, this locus is fully characterized by the solutions of the BPS equations $Q_{\rm loc}\psi_\alpha = 0$. Since we assume that the Weyl multiplet is frozen, we restrict $\alpha$ to label the fermions of the \emph{matter multiplets} only and solve the corresponding BPS equations.

\item {\it Step 4}: Evaluate the classical supergravity action on the localization locus, introducing appropriate boundary counter-terms to cancel the divergent terms arising because of the AdS asymptotics of the near-horizon region.

\item {\it Step 5} (to do in the future):  Determine the one-loop determinant and the integration measure entering the localized QEF and finally perform explicitly the resulting finite-dimensional integral, if possible.

\end{itemize}

Notice that steps 2 through 4 are identical to the usual rigid field theory localization, here on the {\it non-compact} space ${\cal H}_2 \times$S$^2$. Step 1 has only one additional ingredient coming from the supergravity equations of motion, while step 5 is conceptually different due to the fact that Weyl multiplet fields may contribute to the one-loop determinant and may lead to difficulties in defining the integration measure of the localized integral. These issues are more carefully discussed in the concluding section of this paper, while we leave their proper resolution as the main future direction of our work. Given the steps we perform, we can conclude that the progress we report in this paper on the QEF of asymptotically AdS black holes in 4d ${\cal N}=2$ gauged supergravity is similar to the progress originally made in \cite{Dabholkar:2010uh} for asymptotically flat black holes in ungauged supergravity.

\subsection{Main result}
\label{subsec:results}

After performing the procedure described above, we are able to write down the macroscopic black hole degeneracy \eqref{eq:macroscopic-degeneracy} as a constrained finite-dimensional integral\footnote{Note that the exponent corresponds to the classical two-derivative action on the localization locus. Thanks to the conformal formalism, it is straightforward to include higher-derivative terms if needed, which would only result in a change of the integrand but not of the integration domain. When considering such higher-derivative terms, one needs to consider not only chiral superspace integrals as will be done in this paper, but also contributions from full superspace integrals \cite{deWit:2010za}. For asymptotically flat black holes, this was done in \cite{Murthy:2013xpa}.} (see \eqref{eq:dmacro-loc-reg}),
\begin{equation}\label{eq:main-result}
	d_{\rm macro} (p^I, q_I) = \int_{-\infty}^{+\infty} \left( \prod_{I=0}^{n_V} \frac{{\rm d} \phi^I_+}{2\pi} \right) \delta(\xi_I \phi_+^I - 2\pi)\,\exp\Bigl[-2\pi\bigl(p^I\mathcal{F}^+_I(\phi_+) + q_I \phi_+^I\bigr)\Bigr] Z^{\rm reg}_{\mathrm{ind}} (\phi_+) \, .
\end{equation}
Here, $n_V$ is the number of physical vector multiplets and the supergravity theory is uniquely specified by the choice of a holomorphic\footnote{As carefully discussed in later sections, we use the Euclidean formalism where $\phi^I_+$ and $\phi^I_-$ are the analogs of the complex vector multiplet scalars in Lorentzian supergravity.} prepotential $\mathcal{F}^+(\phi_+)$ and constant gauging parameters $\xi_I$. The undetermined part of the above expression, $Z^{\rm reg}_{\mathrm{ind}} (\phi_+)$, comes from the combination of one-loop determinant and integration measure and corresponds to the as-of-yet unfinished step 5 of the localization program outlined above.\footnote{In the case of ungauged supergravity the analog of $Z^{\rm reg}_{\mathrm{ind}}$ has been evaluated explicitly, and it takes a very simple and compact functional form \cite{Murthy:2015yfa,Gupta:2015gga}.} It will be clear that $Z^{\rm reg}_{\mathrm{ind}} (\phi_+)$ only gives $\log(\text{Area})$ and other subleading contributions to the exact black hole entropy. Its explicit form is therefore crucial for the exact evaluation of the above integral but does not alter the saddle-point evaluation, which, as we explicitly show, gives back the correct attractor equations and the expected area dependence of the leading Bekenstein-Hawking entropy. 

We again stress that our gravitational calculation \eqref{eq:main-result}, when evaluated precisely, corresponds to the macroscopic degrees of freedom of the black hole contained inside the event horizon. It is conceivable that the full microscopic degeneracy of the black hole also includes hair degrees of freedom living outside the horizon, as already suggested in  \cite{Liu:2017vll,Liu:2017vbl}. Therefore, the holographic calculation of the microscopic degrees of freedom believed to be captured by the topologically twisted index need not match our macroscopic calculation. Nevertheless we are already able to find some striking agreement with microscopic results. In particular one can already note the similarities between \eqref{eq:main-result} and the formal microcanonical answer for the holographically dual field theory calculation \cite{Benini:2016rke},
\begin{equation}\label{eq:micro-match}
	d_{\rm micro} (\mathfrak{p}_a,\mathfrak{q}_a) = \int_0^{2\pi} \left(\prod_a \frac{\mathrm{d}\Delta_a}{2 \pi} \right) \delta(\sum_a \Delta_a - 2\pi)\ Z(\mathfrak{p}_a, \Delta_a)\ e^{-i \sum_a \Delta_a \mathfrak{q}_a}\ ,
\end{equation}
where $Z(\mathfrak{p}_a, \Delta_a)$ is the field theory partition function depending on the chemical potentials $\Delta_a$ for a set of global $U(1)$ symmetries, and the boundary electromagnetic charges $(\mathfrak{p}_a, \mathfrak{q}_a)$ need to be related to the bulk charges $(p^I, q_I)$ with the proper normalization of the coupling constants dictated by the AdS/CFT correspondence. Note that a leading order match between the integrands of \eqref{eq:main-result} and \eqref{eq:micro-match} has already been established for a number of different holographic cases (see again footnote \ref{footnote1}). We give more details on this already surprising similarity between the macroscopic and the microscopic calculation in several explicit holographic examples in section \ref{sec:examples} and conclude with listing again carefully all the assumptions we made in obtaining \eqref{eq:main-result} and the steps left for future work in section \ref{sec:conclusion}. 

The rest of the paper is devoted to the derivation of \eqref{eq:main-result} as follows. In section \ref{sec:sugra} we give more details on the Euclidean conformal supergravity formalism we use. In section \ref{sec:on-shell} we perform steps 1 and 2 of the localization procedure, presenting the details of the on-shell black hole solution. In section \ref{sec:off-shell} we deal with the off-shell fluctuations characterizing the localization locus of step 3, while in section \ref{sec:loc-action} we perform step 4 by evaluating the classical action on said locus. We leave some of the more technical details in the appendices, where we also gather our conventions.

\section{Euclidean conformal supergravity formalism}
\label{sec:sugra}

Throughout this paper, we make use of the Euclidean 4d $\mathcal{N}=2$ conformal supergravity theory developed in \cite{deWit:2017cle}, and we refer the reader to this paper for more details. This formalism has the advantage of realizing the superconformal algebra $\mathrm{SU}^*(4|2)$ off-shell on the various multiplets. Just as in the Lorentzian case, the theory is equivalent to Poincar\'{e} supergravity upon gauge-fixing the extra conformal (super)symmetries. This gauge-fixing makes use of two compensating multiplets: one vector multiplet, and another multiplet which we choose to be a hypermultiplet. We will therefore consider the superconformal theory describing the Weyl multiplet (containing in particular the graviton) coupled to an arbitrary number $n_V + 1$ of Abelian vector multiplets and a single hypermultiplet (including the compensators).

\subsection{The Weyl multiplet}
\label{subsec:Weyl-sugra}

The Weyl multiplet of the theory comprises the following independent fields
\begin{equation}
\mathbb{W} = (e_\mu{}^a,\,\psi_\mu{}^i,\,b_\mu,\,A_\mu,\,\mathcal{V}_\mu{}^i{}_j,\,T_{ab},\,\chi^i,\,D) \, ,
\end{equation}
as well as composite fields. In the gauge-fixed Poincar\'{e} theory, the field $e_\mu{}^a$ is the vielbein. The rank-2 antisymmetric tensor $T_{ab}$ and the real scalar field $D$ are bosonic auxiliary fields ensuring that the superconformal algebra closes off-shell on this multiplet. Note that, in Euclidean signature, the tensor $T_{ab}$ can be decomposed into self-dual and anti-self-dual components $T_{ab}^\pm$, and both are real and independent\footnote{We use the Euclidean conventions of \cite{deWit:2017cle}, summarized in App. \ref{app:conventions}.}. The other independent bosonic fields are the dilatation gauge field $b_\mu$, the $\mathrm{SO}(1,1)_\mathrm{R}$ gauge field $A_\mu$, and the $\mathrm{SU}(2)_\mathrm{R}$ gauge field $\mathcal{V}_\mu{}^i{}_j$. The fermionic fields are the gravitini $\psi_\mu{}^i$ and an auxiliary fermion, the dilatino, $\chi^i$. All fermions in the theory are symplectic Majorana spinors in Euclidean signature, and can be further decomposed into their chiral left- and right-handed components using the Weyl projectors $(\mathbb{1} \pm \gamma^5)/2$. Note that these chiral components are \emph{not} related by complex conjugation in Euclidean signature. \\

The various transformation rules, as well as the explicit expressions for the composite fields and the weights of the various fields under dilatations and (chiral) $\mathrm{SO}(1,1)_\mathrm{R}$ transformations, can be found in \cite{deWit:2017cle}. For the purpose of our analysis, we will need the transformations for the fermionic fields of the Weyl multiplet under $Q$- and $S$-supersymmetry,
\begin{align}
  \label{eq:gravitino-var}
\delta\psi_\mu{}^i =&\, 2\,\mathcal{D}_\mu\epsilon^i + \tfrac1{16}\mathrm{i}\,T_{ab}\,\gamma^{ab}\,\gamma_\mu\,\epsilon^i - \mathrm{i}\,\gamma_\mu\,\eta^i \, , \\
  \label{eq:dilatino-var}
\delta\chi^i =&\, \tfrac1{24}\mathrm{i}\,\gamma^{ab}\,\Slash{D} T_{ab}\,\epsilon^i + \tfrac16\,R(\mathcal{V})_{ab}{}^i{}_j\,\gamma^{ab}\,\epsilon^j - \tfrac13\,R(A)_{ab}\,\gamma^{ab}\,\gamma^5\,\epsilon^i + D\,\epsilon^i + \tfrac1{24}\,T_{ab}\,\gamma^{ab}\,\eta^i \, ,
\end{align}
where $\epsilon^i$ and $\eta^i$ are the parameters of $Q$- and $S$-supersymmetry, respectively, and $R(\mathcal{V})_{ab}{}^i{}_j$, $R(A)_{ab}$ are the R-symmetry curvatures. The derivative $\mathcal{D}_\mu$ is covariant with respect to all bosonic gauge transformations except special conformal boosts, while $D_\mu$ is fully supercovariant. Note that in a bosonic background these two derivatives coincide, except when acting on $b_\mu$ which is the only independent field that transforms under special conformal boosts. For later reference, we note
\begin{equation}
\mathcal{D}_\mu\epsilon^i = (\partial_\mu - \tfrac14\,\omega_\mu{}^{ab}\,\gamma_{ab} + \tfrac12\,b_\mu + \tfrac12\,A_\mu\,\gamma^5)\,\epsilon^i + \tfrac12\,\mathcal{V}_\mu{}^i{}_j\,\epsilon^j \, ,
\end{equation}
where $\omega_\mu{}^{ab}$ becomes the spin-connection in the gauge-fixed Poincar\'{e} theory.

\subsection{Matter multiplets}
\label{subsec:Matter-sugra}

In addition to the Weyl multiplet, we also consider a number of matter multiplets. The Euclidean vector multiplets involve two real scalar fields\footnote{As opposed to a single complex scalar in Lorentzian signature.} $X_+$ and $X_-$, a symplectic Majorana spinor $\Omega^i$, the gauge field $W_\mu$ and an auxiliary $\mathrm{SU}(2)_\mathrm{R}$ triplet $Y^{ij}$,
\begin{equation}
\mathbb{V}^I = (X^I_+,\,X^I_-,\,W^I_\mu,\,\Omega^{i\,I},\,Y^{ij\,I}) \, ,
\end{equation}
where the index $I$ runs from 0 to $n_V$ (including the compensating multiplet needed to gauge-fix the conformal theory to Poincar\'{e} supergravity). The auxiliary field is subject to the (pseudo-)reality condition $Y_{ij}^I := (Y^{ij\,I})^* = \varepsilon_{ik}\,\varepsilon_{jl}\,Y^{kl\,I}$. The Euclidean transformation rules under $Q$- and $S$-supersymmetry are
\begin{align}
  \label{eq:vector-4D}
\delta X^I_\pm=&\, \pm\mathrm{i}\,\bar\epsilon_{i\pm}\,\Omega^{i\,I}_\pm \, , \nonumber\\[1mm]
\delta W^I_\mu=&\, \bar\epsilon_{i+} \big(\gamma_\mu\,\Omega^{i\,I}_- - 2\mathrm{i}\,X^I_-\psi_\mu{\!}^i{\!}_+ \big) - \bar\epsilon_{i-} \big(\gamma_\mu\,\Omega^{i\,I}_+ - 2\mathrm{i}\,X^I_+\psi_\mu{\!}^i {\!}_-\big) \, , \nonumber\\[1mm]
\delta \Omega^{i\,I}_\pm =&\, -2\mathrm{i}\,\Slash{D} X^I_\pm\,\epsilon^i{\!}_\mp  -\tfrac12\bigl[\hat F(W)_{ab}^{\mp\,I} -\tfrac14 X^I_\mp\,T_{ab}^\mp\bigr]\gamma^{ab} \epsilon^i{\!}_\pm  - \varepsilon_{kj}\, Y^{ik\,I}\epsilon^j{\!}_\pm + 2\,X^I_\pm\,\eta^i{\!}_\pm \, ,  \nonumber   \\[1mm] 
\delta Y^{ij\,I} =&\, 2\,\varepsilon^{k(i}\,\bar{\epsilon}_{k+}\Slash{D}\Omega^{j)\,I}_-  - 2\,\varepsilon^{k(i}\,\bar{\epsilon}_{k-}\Slash{D}\Omega^{j)\,I}_+ \, ,
\end{align}
where the $\pm$ subscripts on the spinors refer to the Weyl projections, while the $\pm$ superscripts on the rank-2 tensors indicate the self-duality. Above, $\hat F(W)_{ab}^I$ denotes the supercovariant field strength of the Abelian gauge field,
\begin{equation}
\hat{F}(W)_{\mu\nu}^I = 2\,\partial_{[\mu} W^I_{\nu]} + \bar{\psi}_{i[\mu}\gamma_{\nu]}\Omega^{i\,I}_+ - \bar{\psi}_{i[\mu}\gamma_{\nu]}\Omega^{i\,I}_- + \mathrm{i}\,X^I_-\bar{\psi}_{\mu\,i}\,\psi_\nu{\!}^i{\!}_+ - \mathrm{i}\,X^I_+\bar{\psi}_{\mu\,i}\,\psi_\nu{\!}^i{\!}_- \, . \vspace{4mm}
\end{equation}

A single Euclidean hypermultiplet contains the following fields
\begin{equation}
\mathbb{H} = (A_i{}^\alpha,\,\zeta^\alpha) \, .
\end{equation}
Here, $A_i{}^\alpha$ are local sections of an $\mathrm{SU}(2)_\mathrm{R}\times\mathrm{Sp}(1)$ bundle, subject to the (pseudo-)reality condition $A^i{}_\alpha := (A_i{}^\alpha)^* = \varepsilon^{ij}\,\Omega_{\alpha\beta}\,A_j{}^\beta$ where $\Omega_{\alpha\beta}$ is a covariantly constant anti-symmetric tensor of~$\mathrm{Sp}(1)$. The spinors $\zeta^\alpha$ are symplectic Majorana according to $C^{-1}\,\bar{\zeta}_\alpha{}^\mathrm{T} = \Omega_{\alpha\beta}\,\zeta^\beta$, with $C$ the charge conjugation matrix. The transformation rules take the form
\begin{align}
  \label{eq:hyper-4D}
\delta A_i{}^\alpha = &\, 2\mathrm{i}\,\bar\epsilon_{i+}\,\zeta^\alpha{\!}_+ - 2\mathrm{i}\,\bar{\epsilon}_{i-}\,\zeta^\alpha{\!}_- \, , \nonumber\\[1mm]
\delta \zeta^\alpha_\pm =&\, -\mathrm{i}\,\Slash{D} A_i{}^\alpha\,\epsilon^i{\!}_\mp - 2\,X^I_\mp\,t_I{}^\alpha{}_\beta\, A_i{}^\beta\epsilon^i{\!}_\pm + A_i{}^\alpha \eta^i{\!}_\pm \, ,
\end{align}
where we have included a coupling to vector multiplets through a gauging described by $n_V + 1$ anti-Hermitian generators $t_I{}^\alpha{}_\beta = \xi_I\,t^\alpha{}_\beta$, with $\xi_I$ the so-called Fayet-Illiopoulos (FI) parameters. The action for hypermultiplets generically takes the form of a supersymmetric sigma model whose target space is a hyper-K\"{a}hler cone, and in this paper we will take the cone to be flat. This implies that the gauging generators~$t_I{}^\alpha{}_\beta$ are constant. The coupling to vector multiplets is also reflected in the covariant derivative of the sections,
\begin{equation}
  \label{eq:section-cov-der}
D_\mu A_i{}^\alpha = (\partial_\mu - b_\mu)A_i{}^\alpha + \tfrac12\,\mathcal{V}_{\mu\,i}{}^j\,A_j{}^\alpha - \xi_I\,W_\mu^I\,t^\alpha{}_\beta\,A_i{}^\beta + \mathrm{fermions} \, ,
\end{equation}
In \eqref{eq:hyper-4D} and \eqref{eq:section-cov-der} we have absorbed the coupling constant $g$ in the FI parameters, so that one recovers the ungauged case by sending all the $\xi_I$ to zero.

\subsection{Off-shell algebra and action}
\label{subsec:Algebra-action-sugra}

The $Q$-supersymmetry transformations close off-shell on the Weyl and vector multiplets, according to the following algebra
\begin{equation}
  \label{eq:QQ}
\big[\delta_\mathrm{Q}(\epsilon_1),\, \delta_\mathrm{Q}(\epsilon_2)\big] = \delta_\mathrm{cgct}(v^\mu) +\delta_\mathrm{M} (\varepsilon_{ab}) + \delta_\mathrm{S} (\hat{\eta}^i) + \delta_\mathrm{K}(\Lambda_\mathrm{K}{\!}^a) + \delta_\mathrm{gauge}(\theta^I) \, ,   
\end{equation}
where $\delta_\mathrm{cgct}$ denotes a covariant general coordinate transformation, while $\delta_\mathrm{M}$, $\delta_\mathrm{S}$ and $\delta_\mathrm{K}$  refer to Lorentz, $S$-supersymmetry and special conformal boost transformations, respectively. The various parameters are
\begin{align}
  \label{eq:parameters-QQ-commutators}
v^\mu =&\, 2\,\bar\epsilon_{2i}\,\gamma^5\gamma^\mu \epsilon_1{}^i \, , \nonumber \\
\varepsilon_{ab} =&\, -\tfrac12\mathrm{i}\,\bar{\epsilon}_{2i-}\,\epsilon_1{\!}^i{\!}_- \, T_{ab}^- + \tfrac12\mathrm{i}\,\bar{\epsilon}_{2i+}\,\epsilon_1{\!}^i{\!}_+ \, T^+_{ab} \, , \\
\hat{\eta}^i =&\, - 6\mathrm{i}\,\bar{\epsilon}_{[1j-} \,\epsilon_{2] }{}^{i}{\!}_-\,\chi^j{\!}_+ + 6\mathrm{i}\,\bar{\epsilon}_{[1j+} \,\epsilon_{2] }{}^{i}{\!}_+\,\chi^j{\!}_- \, , \nonumber \\
\Lambda_\mathrm{K}{\!}^a =&\, - \tfrac12\mathrm{i}\,\bar{\epsilon}_{2i-}\, \epsilon_1{\!}^i{\!}_- \,D_b T^{-ba} + \tfrac12\mathrm{i}\,\bar{\epsilon}_{2i+}\,\epsilon_1{\!}^i{\!}_+\, D_b T^{+ba} + \tfrac32\,\bar{\epsilon}_{2i}\gamma^a\gamma^5\epsilon_1{}^i\, D \, . \nonumber
\end{align}
The gauge transformation $\delta_\mathrm{gauge}$ is associated with the vector multiplet, and has a field-dependent parameter,
\begin{equation}
\label{eq:gauge-param}
\theta^I = 4\mathrm{i}\,\bar{\epsilon}_{2i-}\epsilon_1{}^i{\!}_- X^I_+ - 4\mathrm{i}\,\bar{\epsilon}_{2i+}\epsilon_1{}^i{\!}_+ X^I_- \, .
\end{equation}
We also present the commutator of a $Q$-supersymmetry variation with an $S$-supersymmetry variation, as well as the algebra of $S$-supersymmetry transformations \cite{deWit:2017cle}
\begin{align}
  \label{eq:QS}
\big[\delta_\mathrm{Q}(\epsilon), \, \delta_\mathrm{S}(\eta)\big] =&\; \delta_\mathrm{M}(\mathrm{i}\,\bar\epsilon_i\,\gamma^5\gamma^{ab} \eta^i) + \delta_\mathrm{D}(-\mathrm{i}\,\bar\epsilon_i\,\gamma^5\eta^i) + \delta_{\mathrm{SO}(1,1)} (\mathrm{i}\,\bar\epsilon_i\,\eta^i) \nonumber \\
&\, + \delta_{\mathrm{SU}(2)}(2\mathrm{i}\,\bar\epsilon_j \,\gamma^5\eta^i - \delta_j{\!}^i \,\mathrm{i}\,\bar\epsilon_k\,\gamma^5\eta^k) \, , \\[2mm]
  \label{eq:SS} 
\big[\delta_\mathrm{S}(\eta_1), \, \delta_\mathrm{S}(\eta_2)\big] =&\; \delta_\mathrm{K}(\bar{\eta}_{2i}\,\gamma^5\gamma^a\eta^i_1) \, . \nonumber
\end{align}
It is well-known that on the hypermultiplet the above algebra only closes on-shell, upon using the fermionic equations of motion. Nevertheless, we will discuss in section \ref{subsec:hyper-fluct} how we can close the algebra of a \emph{single} supercharge (which we will take to be the localizing supercharge) off-shell on the hypermultiplet fields.\\

The locally supersymmetric action of the theory under consideration is specified by two prepotentials $\mathcal{F}^\pm(X_\pm)$, which are homogeneous functions of degree 2 in the scalar fields, describing the coupling of the vector multiplets to the Weyl multiplet through a chiral-superspace integral. This construction, as well as the one for hypermultiplet fields transforming under a local gauge group, is presented in \cite{deWit:2017cle}. In App. \ref{app:actions}, we simply collect the expressions that are needed in the following. Here we also note that it is possible to consider more general prepotentials which include a dependence on some independent (anti-)chiral multiplets serving as a background \cite{deWit:2017cle}. Upon identifying such a background with the square of the Weyl multiplet, this generalization leads to higher-derivative terms in the action of Euclidean 4d $\mathcal{N}=2$ supergravity. This is all familiar from the usual Lorentzian supergravity. However, we will restrict ourselves to two-derivative actions in this paper and thus refrain from discussing such a generalization.

\section{On-shell half-BPS attractor}
\label{sec:on-shell}

In this section, we recall the on-shell half-BPS attractor bosonic field configuration for the near-horizon ${\cal H}_2\times$S$^2$ geometry of the black hole solutions in Euclidean gauged 4d $\mathcal{N}=2$ supergravity \cite{deWit:2011gk,Hristov:2016vbm}. This constitutes step 1 of the program outlined in the introduction. As explained there, we will ultimately be interested in \emph{off-shell} fluctuations of matter fields around the on-shell background presented below. Thus, in order to avoid confusion, we will denote the on-shell values of the various matter fields $\Phi$ with a circle, $\mathring{\Phi}$. Since we assume that the Weyl multiplet fields do not fluctuate off-shell, we will not modify the notation for them. After presenting the Euclidean attractor background, we discuss its supersymmetry algebra and conduct step 2 of our program. 

\subsection{The Weyl multiplet}
\label{subsec:Weyl-attractor}

We write the ${\cal H}_2\times$S$^2$ metric in the hyperbolic disk coordinates,
\begin{equation}
\label{eq:1/2-BPS-metric}
ds^2 = v_1\bigl(\sinh^2\eta\,\mathrm{d}\tau^2 + \mathrm{d}\eta^2\bigr) + v_2\bigl(\mathrm{d}\theta^2 + \sin^2\theta\,\mathrm{d}\varphi^2\bigr) \, ,
\end{equation}
where $v_1$ and $v_2$ are real positive constants parameterizing the sizes of the $\mathcal{H}_2$ and $S^2$ spaces, respectively. We use the vielbein one-forms
\begin{equation}
e^1 = \sqrt{v_1}\,\sinh\eta\,\mathrm{d}\tau \, , \;\; e^2 = \sqrt{v_1}\,\mathrm{d}\eta \, , \;\; e^3 = \sqrt{v_2}\,\mathrm{d}\theta \, , \;\; e^4 = \sqrt{v_2}\,\sin\theta\,\mathrm{d}\varphi \, .
\end{equation}
In this coordinate system, the black hole horizon sits at $\eta_H=0$. Another coordinate system we will use is obtained by the change of variable $r:=\cosh\eta$,
\begin{equation}
\label{eq:1/2-BPS-metric-r}
ds^2 = v_1\Bigl((r^2-1)\,\mathrm{d}\tau^2 + \frac{\mathrm{d}r^2}{r^2-1}\Bigr) + v_2\bigl(\mathrm{d}\theta^2 + \sin^2\theta\,\mathrm{d}\varphi^2\bigr) \, .
\end{equation}
In this coordinate system, the horizon of the black hole sits at $r_H=1$.

The non-vanishing components of the auxiliary tensor field in the Weyl multiplet are parameterized by two real scalars $w_\pm$,
\begin{equation}
T_{12}^\mp = -w_\mp \, , \quad T_{34}^\mp = \pm\,w_\mp \, .
\end{equation}
For the half-BPS background, these real scalars are related to the size of ${\cal H}_2$ as \cite{deWit:2011gk}
\begin{equation}
\label{eq:1/2-BPS-v1-w}
v_1^{-1} = -\tfrac14\,w_- w_+ \, .
\end{equation}
The gauge-fixed Poincar\'{e} half-BPS background is obtained by making a gauge choice for the special conformal and $\mathrm{SO}(1,1)_\mathrm{R}$ gauge symmetries. We do so by setting the dilatation and the $\mathrm{SO}(1,1)_\mathrm{R}$ gauge fields to zero, 
\begin{equation}
\label{eq:K-A-gauge-fix}
b_\mu = A_\mu = 0 \, . 
\end{equation}
In addition, we fix the gauge for the local dilatation symmetry by choosing the so-called D-gauge, which sets
\begin{equation}
\label{eq:w-D-gauge-fix}
w_\pm = \pm\,2\,v_1^{-1/2} \, ,
\end{equation}
in accordance with \eqref{eq:1/2-BPS-v1-w}. Note that this gauge choice for the background differs from what is perhaps a more familiar one, where the K\"{a}hler potential (defined in \eqref{eq:Kahler-pot}) is set to a constant. The advantage of using a gauge-fixing of the type \eqref{eq:w-D-gauge-fix} was already explained in \cite{Dabholkar:2010uh}: it allows to keep the $n_V +1$ vector multiplets independent, and uses the degree of freedom associated with the conformal mode of the metric to fix the gauge instead.

The remaining bosonic fields are the auxiliary scalar $D$, which is given by \cite{deWit:2011gk}
\begin{equation}
\label{eq:1/2-BPS-D}
D = -\tfrac16\,\bigl(v_1^{-1} + 2\,v_2^{-1}\bigr) \, ,
\end{equation}
and the $\mathrm{SU}(2)_\mathrm{R}$ gauge field, which we discuss in the next subsection.

\subsection{Matter multiplets}
\label{subsec:Matter-attractor}

As mentioned previously, we will use a single hypermultiplet compensator to ensure that the superconformal theory is gauge-equivalent to Poincar\'{e} supergravity. The sections $A_i{}^\alpha$ of this hypermultiplet can be taken constant on the half-BPS background by an $\mathrm{SU}(2)_\mathrm{R}$ gauge choice,
\begin{equation}
\label{eq:1/2-BPS-hyp}
\chi_\mathrm{H}^{-1/2} \mathring{A}_i{}^\alpha = \delta_i{}^\alpha \, ,
\end{equation}
where the hyper-K\"{a}hler potential is defined as
\begin{equation}
\chi_\mathrm{H} := \tfrac12\,\varepsilon^{ij}\,\Omega_{\alpha\beta}\,\mathring{A}_i^\alpha\,\mathring{A}_j^\beta \, .
\end{equation}
The gauge choice \eqref{eq:1/2-BPS-hyp} breaks the local $\mathrm{SU}(2)_\mathrm{R}$ symmetry to a local $U(1)_\mathrm{R}$ and identifies the indices $\alpha,\beta$ with $i,j$. Hence we can choose, without loss of generality, an explicit diagonal representation for the gauging generators in the transformations \eqref{eq:hyper-4D} \cite{Hristov:2016vbm},
\begin{equation}
\label{eq:gauging-gen}
t^{\,\alpha}{}_\beta = t^{\,i}{}_j := \mathrm{i}\,\sigma_3{}^i{}_j \, ,
\end{equation}
where $\sigma_3$ is the third Pauli matrix. This choice breaks the $\mathrm{SU}(2)_{\rm R}$ invariance of the background down to $U(1)$. It also implies that the so-called moment maps take a simple form on the background:
\begin{equation}
\label{eq:1/2-BPS-moment-maps}
\mu_{ij\,I} := \chi_{\mathrm{H}}^{-1}\,\xi_I\,\mathring{A}_i{}^\alpha\,\Omega_{\alpha\beta}\,t^\beta{}_\gamma\,\mathring{A}_j{}^\gamma = \mathrm{i}\,\xi_I\,\varepsilon_{ik}\,\sigma_3{}^{k}{}_j \, ,
\end{equation}
with $\xi_I$ the FI parameters.\\

We also consider $n_V + 1$ Abelian vector multiplets, including the conformal compensator, coupled to the gravity background. The half-BPS bosonic field configuration satisfies \cite{Hristov:2016vbm}
\begin{equation}
\label{eq:1/2-BPS-vec}
\xi_I\mathring{F}_{34}^{\mp\,I} = \tfrac14\,v_2^{-1} \, , \quad \xi_I\mathring{X}^I_\mp = \tfrac14\,v_1^{-1/2} \, ,
\end{equation}
In addition, the equations of motion for the auxiliary triplet $Y_{ij}^I$ can be derived from the two-derivative action given in \eqref{eq:vector-action} and read
\begin{equation}
\label{eq:Y-EOM-twoderiv}
\mathring{Y}_{ij}^I = 2\,\chi_\mathrm{H}\,\mathring{N}^{IJ}\mu_{ij\,J} \, , 
\end{equation}
where $\mathring{N}^{IJ}$ is the inverse of the matrix 
\begin{equation}
\mathring{N}_{IJ} := \mathcal{F}_{IJ}^+(\mathring{X}_+) + \mathcal{F}_{IJ}^-(\mathring{X}_-) \, ,
\end{equation}
which is built out of derivatives of the prepotentials $\mathcal{F}^\pm$.\\

Let us now come back to the $\mathrm{SU}(2)_\mathrm{R}$ gauge field sitting in the Weyl multiplet. Due to the gauging and supersymmetry, it is related to a linear combination of the gauge fields in the vector multiplets \cite{Hristov:2016vbm}. For the half-BPS background configuration, this imposes
\begin{equation}
\label{eq:1/2-BPS-SU(2)-connection}
\mathcal{V}_\mu{}^i{}_j = -2\mathrm{i}\,\xi_I\mathring{W}_\mu^I\,\sigma_3{}^i{}_j \, .
\end{equation}
In the gauge where $\mathring{W}_\eta^I = 0$, we use the background gauge field components
\begin{equation}
\label{eq:1/2-BPS-vec-connection}
\mathring{W}_\mu^I = \bigl(e^I(\cosh\eta - 1),\,0,\,0,\,-p^I \cos\theta\bigr) \, ,
\end{equation}
where $e^I$ and $p^I$ are the electric fields and the magnetic charges of the black hole solution, respectively.\footnote{Note that the constant included in $\mathring{W}_\tau^I$ is necessary for regularity at the black hole horizon $\eta_H=0$.} In the bulk of this paper, we will consider so-called \emph{electric gaugings}, which means that the linear combinations of electric fields and magnetic charges are fixed,
\begin{equation}
\label{eq:elec-gauging}
\xi_I e^I = 0 \, , \quad \mathrm{and} \quad \, \xi_I p^I = 1/2 \, .
\end{equation}
In turn, \eqref{eq:1/2-BPS-vec-connection} fixes the half-BPS value of the $\mathrm{SU}(2)_\mathrm{R}$ gauge field. In the later stages of the paper, we will briefly comment on more general gaugings. 

For completeness, we note that with the gauging \eqref{eq:elec-gauging}, the non-vanishing components of the field strengths associated with the connections \eqref{eq:1/2-BPS-SU(2)-connection} and \eqref{eq:1/2-BPS-vec-connection} are
\begin{equation}
\mathring{F}^I_{\eta\tau} = e^I \sinh\eta \, , \quad \mathring{F}_{\theta\varphi}^I = p^I \sin\theta \, , \quad R(\mathcal{V})_{\theta\varphi}{}^i{}_j = -\mathrm{i}\sin\theta\,\sigma_3{}^i{}_j \, .
\end{equation}

Let us remark that the Weyl and matter field configuration presented in this section is in agreement with the attractor background expounded in \cite{deWit:2011gk,Hristov:2016vbm}, up to phase factors introduced by our Euclidean formulation.

\subsection{Attractor equations}
\label{subsec:attractor}

An important feature of the half-BPS background presented above are the so-called \emph{attractor equations}. They stem from the (super)symmetry enhancement taking place in the near-horizon geometry of the full black hole solution \cite{Cacciatori:2009iz,Dall'Agata:2010gj,Hristov:2010ri}. In Euclidean signature, the attractor equations are
\begin{equation}
\label{eq:attract}
2\mathrm{i}\,(\mathring{F}_{34}^{\mp\,I} - \tfrac14\,\mathring{X}_\mp^I T_{34}^\mp) - \mathring{Y}^{12\,I} + 4\,\mathring{X}^I_\pm\,\mathring{X}^J_\mp\,\mu_{12\,J} = 0 \, ,
\end{equation}
where the moment maps $\mu_{ij\,I}$ are given in \eqref{eq:1/2-BPS-moment-maps}. Using the equation of motion \eqref{eq:Y-EOM-twoderiv}, this leads to the following relations,
\begin{align}
\label{eq:elec-attract}
v_1\,\mathring{F}_{21}^I =&\; e^I = v_1^{1/2}\,\bigl(\mathring{X}_+^I - \mathring{X}_-^I\bigr) \\
\label{eq:mag-attract}
v_2\,\mathring{F}_{34}^I =&\; p^I = 2\,v_2\,\chi_\mathrm{H}\,\mathring{N}^{IJ}\,\xi_J \, .
\end{align}
These are the electric and magnetic attractor equations, respectively. They will play a central role in the following. \\

In the Euclidean theory, one can also define the following dual field strengths \cite{deWit:2017cle},
\begin{equation}
\label{eq:def-dual}
G^{\mu\nu\pm}{}_I := \pm\frac{1}{e}\,\frac{\partial \mathcal{L}_\mathrm{V}}{\partial F_{\mu\nu}{}^{\pm\,I}} \, ,
\end{equation}
in terms of the Lagrangian density \eqref{eq:vector-action}. With this definition we can compute, at the two-derivative level,
\begin{equation}
\label{eq:attract-dual}
\mathring{G}_{34}{}^\pm{}_I = \pm\bigl[\mathring{\mathcal{F}}_{IJ}^\mp\,\bigl(\mathring{N}^{JK}\chi_\mathrm{H} + 2\,\mathring{X}^J_\mp\,\mathring{X}^K_\pm\bigr)\,\xi_K + \tfrac{1}{2\sqrt{v_1}}\,\mathring{\mathcal{F}}_I^\pm\bigr] \, ,
\end{equation}
where it is understood that $\mathring{\mathcal{F}}^\pm := \mathcal{F}^\pm(\mathring{X}_\pm)$ and similarly for derivatives thereof. 

The attractor equations \eqref{eq:attract} and \eqref{eq:attract-dual} can be written in a manifestly covariant form under the Euclidean electric-magnetic duality group $\mathrm{Sp}(2n_V+2;\,\mathbb{R})$,
\begin{equation}
\label{eq:attract-implicit}
\begin{pmatrix} \mathring{F}_{34}{}^{+\,I} + \mathring{F}_{34}{}^{-\,I} \\ \mathring{G}_{34}{}^+{}_I + \mathring{G}_{34}{}^-{}_I \end{pmatrix} = \frac{1}{v_2}\begin{pmatrix} p^I \\ q_I \end{pmatrix} \, ,
\end{equation}
and
\begin{equation}
\begin{pmatrix} \mathring{F}_{34}{}^{+\,I} - \mathring{F}_{34}{}^{-\,I} \\ \mathring{G}_{34}{}^+{}_I - \mathring{G}_{34}{}^-{}_I \end{pmatrix} = \frac{1}{\sqrt{v_1}}\begin{pmatrix} \mathring{X}^I_- - \mathring{X}^I_+ \\ \mathring{\mathcal{F}}_I^- + \mathring{\mathcal{F}}_I^+ \end{pmatrix} + \chi_\mathrm{H}\,\begin{pmatrix} 0 \\ \xi_I \end{pmatrix} \, , \vspace{2mm}
\end{equation}
where we have introduced the electric charges $q_I$ of the black hole solution. Equation \eqref{eq:attract-implicit} together with \eqref{eq:attract-dual} also implies the useful on-shell relation
\begin{equation}
\label{eq:q-p-attract}
q_I = \frac12\,(\mathring{\mathcal{F}}_{IJ}^- - \mathring{\mathcal{F}}_{IJ}^+)\,p^J \, .
\end{equation}

\subsection{Supersymmetry algebra and the localizing supercharge}
\label{subsec:KS}

We now turn to the Euclidean (conformal) Killing spinors associated with the attractor geometry. These are spinor parameters for which the supersymmetry variations of the gravitino and the auxiliary dilatino sitting in the Weyl multiplet, \eqref{eq:gravitino-var} and \eqref{eq:dilatino-var}, vanish. We relegate the details of their derivation and their explicit expressions to App.~\ref{app:CKS}. There, we show that the on-shell half-BPS attractor background admits two complex unbroken conformal supercharges which we parameterize by \emph{commuting} spinor parameters $(\xi^i,\kappa^i)$ and $(\widetilde{\xi}^{\,i},\widetilde{\kappa}^{\,i})$, $\mathcal{Q} := (\xi^i)^\dagger\,Q^i + (\kappa^i)^\dagger\,S^i$ and $\widetilde{\mathcal{Q}} := (\widetilde{\xi}^i)^\dagger\,Q^i + (\widetilde{\kappa}^i)^\dagger\,S^i$. 

We can evaluate the algebra of the unbroken supercharges starting from the off-shell algebra \eqref{eq:QQ} and \eqref{eq:QS}. The result is 
\begin{align}
\label{eq:loc-alg-QQ}
\mathcal{Q}^{\,2} =&\; \delta_{\mathrm{cgct}}(\mathring{v}^\mu) + \delta_\mathrm{M}(\mathring{\varepsilon}_{ab}) + \delta_{\mathrm{SU}(2)}(\mathring{\Lambda}_j{}^i) + \delta_{\mathrm{K}}(\mathring{\Lambda}_K{}^a) + \delta_\mathrm{gauge} \, , \\
\label{eq:loc-alg-QtQt}
\widetilde{\mathcal{Q}}^{\,2} =&\; \delta_{\mathrm{cgct}}(\mathring{\widetilde{v}}{}^\mu) + \delta_\mathrm{M}(\mathring{\varepsilon}_{ab}) + \delta_{\mathrm{SU}(2)}(\mathring{\Lambda}_j{}^i) + \delta_{\mathrm{K}}(\mathring{\widetilde{\Lambda}}_K{}^a) + \delta_\mathrm{gauge} \, , \\
\label{eq:loc-alg-QQt}
\bigl\{\mathcal{Q},\,\widetilde{\mathcal{Q}}\,\bigr\} =&\; \delta_\mathrm{M}(\mathring{\lambda}_{ab}) + \delta_\mathrm{gauge} \, ,
\end{align}
and the parameters on the r.h.s. are as follows,
\begin{align}
\mathring{v}^\mu =&\; - \mathring{\widetilde{v}}{}^\mu = \frac{2}{\sqrt{v_1}} \begin{pmatrix} |\alpha|^2 + |\beta|^2 - \mathrm{i}e^{-\mathrm{i}\tau}\bar{\alpha}\beta\coth\eta + \mathrm{i}e^{\mathrm{i}\tau}\alpha\bar{\beta}\coth\eta \\ e^{-\mathrm{i}\tau}\bar{\alpha}\beta + e^{\mathrm{i}\tau}\alpha\bar{\beta} \\ 0 \\ 0 \end{pmatrix} \, , \\[1mm]
\mathring{\varepsilon}_{12} =&\; -\frac{2}{\sqrt{v_1}}\Bigl((|\alpha|^2 + |\beta|^2)\cosh\eta - \mathrm{i}e^{-\mathrm{i}\tau}\bar{\alpha}\beta\sinh\eta + \mathrm{i}e^{\mathrm{i}\tau}\alpha\bar{\beta}\sinh\eta\Bigr) \, , \\[1mm]
\mathring{\Lambda}_j{}^i =&\; -\frac{|\alpha|^2 - |\beta|^2}{\sqrt{v_1}}\,\mathrm{i}\,\sigma_3{}^i{}_j \, , \\[1mm]
\mathring{\Lambda}_K{}^a =&\; - \mathring{\widetilde{\Lambda}}_K{}^a = \frac{v_1 + v_2}{4 v_1 v_2}\,\mathring{v}^a \, , \\[1mm]
\mathring{\lambda}_{12} =&\; -\frac{4}{\sqrt{v_1}}(|\alpha|^2 - |\beta|^2) \, ,
\end{align}
where $\alpha$ and $\beta$ are arbitrary complex constants parameterizing the unbroken conformal supercharges (see App.~\ref{app:CKS} for details). The special conformal boost transformation and the gauge transformation with field-dependent parameter will not be relevant in what follows, so we refrain from giving their explicit parameters. We do expect the gauge transformation to play a central role \cite{Murthy:2015yfa,Gupta:2015gga} in the computation of the one-loop determinant (step 5 of our program) for matter fields around the off-shell field configuration presented in the next section.\\

Moving on to step 2 of the localization program in the introduction, we choose the localizing supercharge $Q_{\rm loc}$ parameterized by $(\xi^i,\kappa^i)$ with $\beta \in \mathbb{R}^*$ and $\alpha=0$. It squares to the ${\cal H}_2$ isometry $L_0 \sim \mathrm{i}\,\partial_\tau$ and additional internal symmetries, which is consistent with the general isometry algebra of the near-horizon geometry presented in App. \ref{app:superalgebra}. Furthermore, by splitting off the Lorentz and special conformal transformation parts of the covariant general coordinate transformation on the r.h.s. of \eqref{eq:loc-alg-QQ} and combining them with the Lorentz and special conformal transformations parameterized by $\mathring{\varepsilon}_{ab}$ and $\mathring{\Lambda}_K{}^a$, respectively, we can write the algebra of our localizing supercharge as
\begin{equation}
\label{eq:loc-algebra}
Q_{\rm loc}{}^2 = \mathcal{L}_{\mathring{v}} + \delta_{\mathrm{SU}(2)}\bigl(\tfrac{1}{\sqrt{v_1}}\,\mathrm{i}\,\sigma_3{}^i{}_j\bigr) + \delta_\mathrm{gauge} \, ,
\end{equation}
with $\mathcal{L}_{\mathring{v}}$ is the usual Lie derivative along the vector $\mathring{v}^\mu = \tfrac{2}{\sqrt{v_1}}(1,0,0,0)^{\rm T}$. Observe that the Lorentz and special conformal transformations cancel due to our choice of gauge for the local Lorentz and special conformal symmetries, in which
\begin{equation}
\mathring{\varepsilon}_{ab} - \mathring{v}^\mu\,\omega_{\mu\,ab} = 0 \, , \quad \textnormal{and} \quad \mathring{\Lambda}_K{}^a - \mathring{v}^\mu\,f_\mu{}^a = 0 \, .
\end{equation}
Aside from the field-dependent gauge transformation inherent to (conformal) supergravity, one should note the similarities between \eqref{eq:loc-algebra} and the algebra of the localizing supercharge used for the computation of the topologically twisted index in the dual $\mathrm{CFT}_3$ \cite{Benini:2015noa}. Evidently, these similarities are in no small part due to our working hypothesis that the Weyl multiplet, and in turn the parameters $(\xi^i,\kappa^i)$, are frozen to their on-shell half-BPS values. 

\section{Off-shell fluctuations}
\label{sec:off-shell}

Having presented the details of the on-shell half-BPS attractor configuration, we now turn to analyzing \emph{off-shell} bosonic fluctuations around this background which annihilate the $Q_{\rm loc}$-transformations of the various fermions in the theory. Thanks to our conformal supergravity set-up, we can do so in each multiplet separately. The first step would be to analyze the Weyl multiplet, but as we stressed in the introduction, we work under the assumption that the Weyl multiplet fields are pinned to their on-shell values. We remind the reader that this was indeed proven to be true for BPS black holes in ungauged supergravity \cite{Gupta:2012cy}. In addition, the consistency of our final result \eqref{eq:main-result} with previous results lends a certain amount of confidence in this assumption. We thus move on and consider the bosonic fluctuations around the half-BPS background for the matter multiplet fields.   

\subsection{Vector multiplets}
\label{subsec:vector-fluct}

In a bosonic background, the action of the localizing supercharge on the gaugino of a given vector multiplet with index $I=0, \ldots, n_V$ follows from \eqref{eq:vector-4D},
\begin{equation}
\label{eq:gaugino-var}
Q_{\rm loc}\,\Omega^{i\,I}_\pm = -2\mathrm{i}\,\Slash{\mathcal{D}}X^I_\pm\,\xi^i_\mp - \tfrac12\bigl[F_{ab}^{\mp\,I} - \tfrac14\,X^I_\mp T_{ab}^\mp\,\bigr]\gamma^{ab}\xi^i_\pm - \varepsilon_{kj}\,Y^{ik\,I}\xi^j_\pm + 2\,X^I_\pm\,\kappa^i_\pm \, ,
\end{equation}
where $\xi^i_\pm$ and $\kappa^i_\pm$ are given in \eqref{eq:xi-1}, \eqref{eq:kappa-1}, \eqref{eq:xi-2} and \eqref{eq:kappa-2}, with $\beta \in \mathbb{R}^*$ and $\alpha=0$ as discussed in the previous section. Closing the variation \eqref{eq:gaugino-var} from the left with $\bar{\xi}_{j\pm}$ and $\bar{\kappa}_{j\pm}$, we obtain\footnote{One can check that closing with $\bar{\xi}_j\Gamma$ and $\bar{\kappa}_j\Gamma$, where $\Gamma$ is a rank-1 or rank-2 element of the Clifford algebra, does not yield independent equations.} the following independent equations (note that each equation below encodes two equations, one for each sign):
\begin{align}
\label{eq:vec-0-1}
0 =&\; 2\,\sinh\eta\,(\mathrm{i}\,\partial_3 + \partial_4)X^I_\pm + (\cosh\eta \mp 1)\,e^{-\mathrm{i}\tau}\,Y^{22\,I} \, , \\
\label{eq:vec-0-2}
0 =&\; 2\,\sinh\eta\,(\mathrm{i}\,\partial_2 \pm\partial_1)X^I_\pm \pm \mathrm{i}\,(\cosh\eta \mp 1)\Bigl[F_{12}^{\mp\,I} \mp F_{34}^{\mp\,I} +\tfrac{1}{\sqrt{v_1}}(X^I_+ - X^I_-) \mp \mathrm{i}\,Y^{12\,I}\Bigr] \, ,
\end{align}
Using the Euclidean reality conditions
\begin{equation}
\label{eq:Euclidean-reality}
(X^I_\pm)^\dagger = X^I_\pm \, , \quad (F^I_{ab})^\dagger = F^I_{ab} \, , \quad (Y^{ij\,I})^\dagger = \varepsilon_{ik}\varepsilon_{jl}\,Y^{kl\,I} \, , 
\end{equation}
we can separate these complex equations into their real and imaginary parts:
\begin{align}
\label{eq:vec-0-1-Re}
0 =&\; 2\,\sinh\eta\,\partial_4 X^I_\pm + (\cosh\eta \mp 1)\,\mathrm{Re}\bigl[e^{-\mathrm{i}\tau}\,Y^{22\,I}\bigr] \, , \\
\label{eq:vec-0-1-Im}
0 =&\; 2\,\sinh\eta\,\partial_3 X^I_\pm + (\cosh\eta \mp 1)\,\mathrm{Im}\bigl[e^{-\mathrm{i}\tau}\,Y^{22\,I}\bigr] \, , \\
\label{eq:vec-0-2-Re}
0 =&\; 2\,\sinh\eta\,\partial_1 X^I_\pm \, , \\
\label{eq:vec-0-2-Im}
0 =&\; 2\,\sinh\eta\,\partial_2 X^I_\pm \pm (\cosh\eta \mp 1)\Bigl[F_{12}^{\mp\,I} \mp F_{34}^{\mp\,I} +\tfrac{1}{\sqrt{v_1}}(X^I_+ - X^I_-) \mp \mathrm{i}\,Y^{12\,I}\Bigr] \, .
\end{align}

We now linearly split the bosonic fields into the sum of their background on-shell values and off-shell fluctuations, 
\begin{equation}
\label{eq:vec-background-split}
X_\pm^I = \mathring{X}_\pm^I + \frac{1}{2}\bigl(\Sigma^I_+ \pm \Sigma^I_-\bigr) \, , \quad Y^{ij\,I} = \mathring{Y}^{ij\,I} + y^{ij\,I} \, , \quad F_{ab}^I = \mathring{F}_{ab}^I + f_{ab}^I \, ,
\end{equation}
where the fluctuations $\Sigma_\pm$, $y^{ij}$ and $f_{ab}$ are (pseudo-)real owing to the reality conditions \eqref{eq:Euclidean-reality}. The on-shell field configuration solves \eqref{eq:vec-0-1-Re}, \eqref{eq:vec-0-1-Im}, \eqref{eq:vec-0-2-Re} and \eqref{eq:vec-0-2-Im}, and these equations are linear in the fields so we are left with the following BPS equations on the fluctuations:
\begin{align}
\label{eq:vec-fluct-0-1-Re}
0 =&\; \sinh\eta\,\partial_4\bigl(\Sigma^I_+ \pm \Sigma^I_-\bigr) + (\cosh\eta \mp 1)\,\mathrm{Re}\bigl[e^{-\mathrm{i}\tau}\,y^{22\,I}\bigr] \, , \\
\label{eq:vec-fluct-0-1-Im}
0 =&\; \sinh\eta\,\partial_3\bigl(\Sigma^I_+ \pm \Sigma^I_-\bigr) + (\cosh\eta \mp 1)\,\mathrm{Im}\bigl[e^{-\mathrm{i}\tau}\,y^{22\,I}\bigr] \, , \\
\label{eq:vec-fluct-0-2-Re}
0 =&\; \sinh\eta\,\partial_1\bigl(\Sigma^I_+ \pm \Sigma^I_-\bigr) \, , \\
\label{eq:vec-fluct-0-2-Im}
0 =&\; \sinh\eta\,\partial_2\bigl(\Sigma^I_+ \pm \Sigma^I_-\bigr) \pm (\cosh\eta \mp 1)\Bigl[f^I_{12} \mp f^I_{34} +\tfrac{1}{\sqrt{v_1}}\Sigma^I_- \mp \mathrm{i}\,y^{12\,I}\Bigr] \, .
\end{align}
Taking a trivial linear combination of \eqref{eq:vec-fluct-0-2-Re} leads to
\begin{equation}
\label{eq:vec-fluct-tau-ind}
\partial_\tau \Sigma^I_\pm = 0 \, .
\end{equation}
Using this and taking $\tau$-derivatives of the other equations leads to $\partial_\tau(e^{-\mathrm{i}\tau}\,y^{22\,I}) = 0$, which is solved for
\begin{equation}
y^{22\,I} = e^{\mathrm{i}\tau}\,(y^I_\mathfrak{R} + \mathrm{i}\,y^I_\mathfrak{I}) \, ,
\end{equation}
where~$y^I_\mathfrak{R}$ and $y^I_\mathfrak{I}$ are real functions of $(\eta,\theta,\varphi)$. All other quantities are $\tau$-independent. We can now solve for $y^I_\mathfrak{R}$ and $y^I_\mathfrak{I}$ in terms of angular derivatives on $\Sigma^I_-$ by subtracting the upper and lower sign equations in \eqref{eq:vec-fluct-0-1-Re} and \eqref{eq:vec-fluct-0-1-Im}: 
\begin{equation}
\label{eq:vec-fluct-y11}
y^I_\mathfrak{R} = \sinh\eta\,\partial_4\Sigma_-^I \, , \quad y^I_\mathfrak{I} = \sinh\eta\,\partial_3 \Sigma_-^I \, .
\end{equation}
Using this in the sum of the upper and lower sign equations in \eqref{eq:vec-fluct-0-1-Re} and \eqref{eq:vec-fluct-0-1-Im}, we also obtain the following equation for the angular derivatives of $\Sigma_\pm^I$,
\begin{align}
\label{eq:vec-BPS-1}
\sinh\eta\,\partial_3\bigl[\cosh\eta\,\Sigma^I_- + \Sigma^I_+\bigr] =&\; 0 \, , \\
\label{eq:vec-BPS-2}
\sinh\eta\,\partial_4\bigl[\cosh\eta\,\Sigma^I_- + \Sigma^I_+\bigr] =&\; 0 \, .
\end{align}

It remains to analyze \eqref{eq:vec-fluct-0-2-Im}, which encodes a system of two first-order differential equations for the $\eta$ dependence of $\Sigma_\pm^I$. Taking linear combinations, one can show that it is equivalent to the system
\begin{align}
\label{eq:vec-fluct-0-2-1}
0 =&\; \sinh\eta\,\partial_2\Sigma_+^I - \tfrac{1}{\sqrt{v_1}}\,\Sigma_-^I - f_{12}^I - \cosh\eta\,f_{34}^I - \mathrm{i}\,\cosh\eta\,y^{12\,I} \, , \\
\label{eq:vec-fluct-0-2-2}
0 =&\; \sinh\eta\,\partial_2\Sigma_-^I + \tfrac{1}{\sqrt{v_1}}\cosh\eta\,\Sigma_-^I + \cosh\eta\,f_{12}^I + f_{34}^I + \mathrm{i}\,y^{12\,I} \, .
\end{align}
We immediately impose $f_{34}^I = 0$ since the magnetic charges of the black hole are fixed at the quantum level and cannot fluctuate off-shell.  Furthermore, the fluctuations of the field strength are constrained due to the gauging \eqref{eq:1/2-BPS-SU(2)-connection}, which implies 
\begin{equation}
\label{eq:fluct-f-constraint}
\xi_I f_{12}^I = 0 \, .
\end{equation}
Multiplying \eqref{eq:vec-fluct-0-2-2} by $\cosh\eta$ and adding \eqref{eq:vec-fluct-0-2-1} yields
\begin{equation}
\label{eq:vec-BPS-3}
\sinh\eta\,\partial_2\bigl[\cosh\eta\,\Sigma^I_- + \Sigma_+^I\bigr] + \sinh^2\eta\,f_{12}^I = 0 \, ,
\end{equation}
while multiplying \eqref{eq:vec-fluct-0-2-1} by $\cosh\eta$ and adding \eqref{eq:vec-fluct-0-2-2} yields
\begin{equation}
\label{eq:vec-BPS-4}
\sinh\eta\,\partial_2\bigl[\cosh\eta\,\Sigma^I_+ + \Sigma_-^I\bigr] - \sinh^2\eta\,\Bigl(\tfrac{1}{\sqrt{v_1}}\,\Sigma_+^I + \mathrm{i}\,y^{12\,I}\Bigr) = 0 \, .
\end{equation}

We look for smooth and regular solutions to the BPS equations \eqref{eq:vec-BPS-1}, \eqref{eq:vec-BPS-2}, \eqref{eq:vec-BPS-3} and \eqref{eq:vec-BPS-4}. These solutions should furthermore respect the boundary and fall-off conditions \cite{Castro:2008ms} imposed by the QEF \eqref{eq:macroscopic-degeneracy}. This implies that we can take the following Ansatz\footnote{Note that there is no term proportional to $\cosh^{-1}\eta$ in $f^I(\eta)$, as such a term would imply a term proportional to $\log(\cosh\eta) = \log r$ in the gauge fields, with $r$ the radial coordinate on $\mathcal{H}_2$. This is forbidden by the fall-off conditions \cite{Castro:2008ms}.},
\begin{align}
\label{eq:fluct-ansatz}
\Sigma_-^I =&\; \sum_{k=1}^{\infty} \frac{C^I_k(\theta,\varphi)}{\cosh^k\eta} \, , \qquad\quad \Sigma_+^I = \sum_{k=1}^{\infty} \frac{D^I_k(\theta,\varphi)}{\cosh^k\eta} \, , \\
y^{12\,I} =&\, \frac{\mathrm{i}}{\sqrt{v_1}}\,\sum_{k=1}^{\infty} \frac{y^I_k(\theta,\varphi)}{\cosh^k\eta} \, , \quad f_{12}^I = \frac{1}{\sqrt{v_1}}\,\sum_{k=2}^{\infty} \frac{f^I_k(\theta,\varphi)}{\cosh^k\eta} \, ,
\end{align}
where $C_k^I(\theta,\varphi)$, $D_k^I(\theta,\varphi)$, $y_k^I(\theta,\varphi)$ and $f_k^I(\theta,\varphi)$ are real functions of the angular coordinates and parameterize the fluctuations. In what follows, we will often suppress the angular dependence to lighten the notation.
 
The BPS equations \eqref{eq:vec-BPS-3} and \eqref{eq:vec-BPS-4} provide us with relations between the various coefficients when analyzed order-by-order in $\cosh^{-1}\eta$, or equivalently in $1/r$ in the coordinate system \eqref{eq:1/2-BPS-metric-r} for $\mathcal{H}_2$. We obtain, for all $k\geq 1$,
\begin{align}
\label{eq:vec-BPS-f-gen}
\sinh^2\eta\,\bigl[(k-1)(C^I_k + D_{k-1}^I) - f_k^I\,\bigr] =&\; 0 \, , \\
\label{eq:vec-BPS-y-gen}
\sinh^2\eta\,\bigl[(k-1)\,C^I_{k-1} + k\,D_{k}^I - y_k^I\,\bigr] =&\; 0 \, ,
\end{align}
where it is understood that we set $f_1^I = D_0^I := 0$. In the same way, \eqref{eq:vec-BPS-1} and \eqref{eq:vec-BPS-2} imply, for all $k\geq 1$,
\begin{equation}
\label{eq:vec-BPS-ang-gen}
\sinh\eta\,\partial_3\bigl[C^I_k + D_{k-1}^I\bigr] = \sinh\eta\,\partial_4\bigl[C^I_k + D_{k-1}^I\bigr] = 0 \, 
\end{equation}
Away from the horizon $\eta = 0$, \eqref{eq:vec-BPS-f-gen} and \eqref{eq:vec-BPS-y-gen} are solved by
\begin{align}
\label{eq:vec-BPS-f-nohor}
f_k^I = (k-1)(C_k^I + D_{k-1}^I) \, , \\
\label{eq:vec-BPS-y-nohor}
y^I_k =(k-1)\,C_{k-1}^I + k\,D_k^I \, , 
\end{align}
for all $k\geq 1$, and \eqref{eq:vec-BPS-ang-gen} shows that $\partial_3 f_k^I = \partial_4 f_k^I = 0$, which in turn means that the field strength fluctuations are constant on the $S^2$,
\begin{equation}
\label{eq:f-S2-ind}
\partial_3 f_{ab}^I = \partial_4 f_{ab}^I = 0 \, .
\end{equation}
At the horizon $\eta=0$, the BPS equations degenerate because some of the Killing spinor parameters vanish (see App. \ref{app:CKS}). However, the above BPS configuration is also a solution of \eqref{eq:vec-fluct-0-2-1} and \eqref{eq:vec-fluct-0-2-2} at the horizon and there is no topological obstruction to continue them at the point $\eta=0$. The result \eqref{eq:f-S2-ind} then shows that the field strength fluctuations are constant on the 2-sphere at every point in $\mathcal{H}_2$.

We stress that the BPS configuration derived in this section is non-singular. Were we to allow for singular configurations, there could be additional contributions to the localization manifold. We will not consider this case in this paper, although we do remind the reader that, in the analysis of the QEF for 1/2-BPS black holes in ungauged supergravity, there were exponentially suppressed contributions coming from stringy orbifolds of the near-horizon geometry of the form AdS$_2/\mathbb{Z}_c$, with $c>1$ \cite{Dabholkar:2014ema}. We leave the parallel analysis in gauged supergravity for future work.

Finally, we note that the constraint on the gauge field fluctuations \eqref{eq:fluct-f-constraint} can be written as a constraint on the free parameters $C_k^I$ and $D_k^I$,
\begin{equation}
\label{eq:fluct-f-constraint-CD}
\xi_I(C_{k+1}^I + D_k^I) = 0 \;\; \textnormal{for all} \;\; k \geq 1 \, .
\end{equation}

\subsection{The compensating hypermultiplet}
\label{subsec:hyper-fluct}

We now discuss off-shell fluctuations of the compensating hypermultiplet. As is well-known, the full superconformal algebra only closes off-shell on hypermultiplets after the introduction of an infinite tower of auxiliary fields. There is, however, a technique to close the algebra of a \emph{single} supercharge (here the localizing supercharge $Q_{\rm loc}$) off-shell \cite{Berkovits:1993hx}. To do so, we introduce a set of 4 (pseudo-)real auxiliary fields $H_i{}^\alpha$ and find the appropriate transformation rules so as to close the algebra of the localizing supercharge off-shell. Explicitly, the action of $Q_{\rm loc}$ in a bosonic background on the hyperino $\zeta^\alpha_\pm$ and the auxiliary fields are modified compared to their on-shell formulation \eqref{eq:hyper-4D} as follows
\begin{align}
\label{eq:hyperino-var}
Q_{\rm loc}\,\zeta^\alpha_\pm =&\; -\mathrm{i}\Slash{\mathcal{D}}A_i{}^\alpha \xi^i_\mp - 2\,\xi_I X^I_\mp\,t^\alpha{}_\beta A_i{}^\beta \xi^i_\pm + H_i{}^\alpha\,\check{\xi}^i_\pm + A_i{}^\alpha \kappa^i_\pm \, , \\
Q_{\rm loc}\,H_i{}^\alpha =&\; \bar{\check{\xi}}_{i+}\Slash{\mathcal{D}}\zeta^\alpha_- - \bar{\check{\xi}}_{i-}\Slash{\mathcal{D}}\zeta^\alpha_+ \, .
\end{align}
where we introduced \emph{constrained} symplectic Majorana-Weyl parameters $\check{\xi}^i_\pm$. Note that the on-shell case is recovered by setting the auxiliary scalars to zero. The localizing supercharge $Q_{\rm loc}$ closes according to \eqref{eq:loc-algebra} without the need to use fermionic equations of motion provided the constrained parameters satisfy
\begin{align}
\bar{\xi}_{i+}\check{\xi}^j_+ =&\; \bar{\xi}_{i-}\check{\xi}^j_- \, , \nonumber \\
\bar{\check{\xi}}_{i\pm}\check{\xi}^j_\pm =&\; \bar{\xi}_{i\mp}\xi^j_\mp \, , \\
\bar{\check{\xi}}_{i\pm}\gamma^\mu\check{\xi}^j_\mp =&\; \bar{\xi}_{i\pm}\gamma^\mu\xi^j_\mp \, . \nonumber
\end{align}
These constraints have a non-trivial solution for the parameter $\xi^i$ of $Q_{\rm loc}$, given by
\begin{equation}
\label{eq:xi-check}
\check{\xi}^i_+ = \Bigl(\frac{\cosh\eta - 1}{\cosh\eta + 1}\Bigr)^{-1/2}\,\xi^i_+\ , \quad \quad \check{\xi}^i_- = \Bigl(\frac{\cosh\eta - 1}{\cosh\eta + 1}\Bigr)^{1/2}\,\xi^i_- \, .
\end{equation}

Since the Weyl multiplet and in particular the $\mathrm{SU}(2)_\mathrm{R}$ connection are fixed to their on-shell values, the hypermultiplet conformal compensator should fix the $\mathrm{SU}(2)_\mathrm{R}$ gauge not just in the attractor background but at the full off-shell level. This implies that we can take the hyper sections $A_i{}^\alpha = \mathring{A}_i{}^\alpha + a_i{}^\alpha$ such that
\begin{equation}
\label{eq:quantum-SU2-gauge}
\chi_\mathrm{H}^{-1/2} A_i{}^\alpha = \delta_i{}^\alpha \, ,
\end{equation}
which, according to \eqref{eq:1/2-BPS-hyp}, fixes $a_i{}^\alpha = 0$. Equivalently, the choice \eqref{eq:quantum-SU2-gauge} preserves the identification of the $\mathrm{Sp}(1)$ bundle with $\mathrm{SU}(2)_\mathrm{R}$ for the off-shell fluctuations, and thus does not modify our explicit representation of the gauging generators $t^\alpha{}_\beta$ \eqref{eq:gauging-gen}.

Just as in the vector multiplet case, \eqref{eq:hyperino-var} is solved for the on-shell half-BPS values of the various fields. Because of the gauging, we are thus left with an equation relating the auxiliary fields $H_i{}^\alpha$ and the fluctuations of the vector multiplet scalars analyzed in the previous subsection. Explicitly, we have
\begin{equation}
H_i{}^\alpha\,\check{\xi}^i_\pm - \xi_I (\Sigma^I_+ \mp \Sigma^I_-)\,t^\alpha{}_i\,\xi^i_\pm = 0 \, .
\end{equation}
Using the constrained parameters \eqref{eq:xi-check} and $t^i{}_j = \mathrm{i}\,\sigma_3{}^i{}_j$, this is
\begin{equation}
\Bigl[H_i{}^\alpha\,\Bigl(\frac{\cosh\eta - 1}{\cosh\eta + 1}\Bigr)^{\mp1/2} - \mathrm{i}\,\xi_I (\Sigma^I_+ \mp \Sigma^I_-)\,\sigma_3{}^\alpha{}_i\Bigr]\xi^i_\pm = 0 \, ,
\end{equation}
which is solved for
\begin{equation}
\label{eq:hyper-aux}
H_i{}^\alpha = \mathrm{i}\,\sigma_3{}^\alpha{}_i\,\Bigl(\frac{\cosh\eta - 1}{\cosh\eta + 1}\Bigr)^{\pm1/2}\,\xi_I (\Sigma^I_+ \mp \Sigma^I_-) \, .
\end{equation}

We now observe that these two equations immediately imply
\begin{equation} 
\xi_I \Sigma^I_+ + \cosh\eta\,\xi_I \Sigma^I_- = 0 \, .
\end{equation}
Using the expressions \eqref{eq:fluct-ansatz}, this yields
\begin{equation}
\label{eq:constraint-hypers}
\xi_I C^I_1 = 0 \, , \quad \quad \xi_I(C_{k+1}^I + D_k^I) = 0\ , \;\; \quad \textnormal{for all} \;\; k \geq 1 \, .
\end{equation}
With the second equation above, we recover precisely the condition $\xi_If_{12}^I = 0$ \eqref{eq:fluct-f-constraint-CD}, which is a consistency check on our procedure.

\subsection{Summary of the localizing manifold}
\label{subsec:loc-manifold}

In this section, we have achieved step 3 of our localization program by characterizing the localization locus for $n_V+1$ Abelian vector multiplets and the compensating hypermultiplet. We summarize the results here for convenience. In the coordinate system \eqref{eq:1/2-BPS-metric-r} for $\mathcal{H}_2$, the off-shell BPS fluctuations of the matter fields are controlled by two functions,
\begin{equation}
\Sigma_-^I = \sum_{k=1}^{\infty} \frac{C^I_k(\theta,\varphi)}{r^k} \, , \qquad \Sigma_+^I = \sum_{k=1}^{\infty} \frac{D^I_k(\theta,\varphi)}{r^k} \, ,
\end{equation}
with real functions $C_k^I(\theta,\varphi)$ and $D_k^I(\theta,\varphi)$ subject to the constraints
\begin{align}
\label{eq:full-loc-C1-constraint}
&\xi_I C^I_1 = 0 \, , \\
\label{eq:full-loc-Ck-constraint}
&\xi_I(C_{k+1}^I + D_k^I) = 0 \;\; \textnormal{for all} \;\; k \geq 1 \, .
\end{align}
In terms of these functions, the bosonic field configuration which solves the BPS equations with respect to the localizing supercharge $Q_{\rm loc}$ is as follows.\footnote{Here we include the on-shell values derived in section \ref{sec:on-shell}.}\\
\underline{Vector multiplets}:
\begin{align}
\label{eq:full-loc-vectors}
X^I_\pm =&\, \mathring{X}^I_\pm + \frac12\,\bigl(\Sigma_+^I \pm \Sigma_-^I\bigr) \, , \nonumber \\
F_{21}^I =&\, \frac{e^I}{v_1} + \frac{1}{\sqrt{v_1}}\bigl(\Sigma^I_- + r\,\partial_r\Sigma^I_- + \partial_r\Sigma^I_+\bigr) \, , \quad F_{34}^I = \frac{p^I}{v_2} \, , \\
Y^{12\,I} =&\, \mathrm{i}\,\frac{p^I}{v_2} - \frac{\mathrm{i}}{\sqrt{v_1}}\bigl(r\,\partial_r\Sigma^I_+ + \partial_r\Sigma_-^I\bigr) \, , \nonumber \\
Y^{22\,I} =&\; (Y^{11\,I})^\dagger = e^{\mathrm{i}\tau}\,\sqrt{r^2-1}\,(\partial_4 + \mathrm{i}\,\partial_3)\Sigma_-^I \, . \nonumber
\end{align}
\underline{Compensating hypermultiplet}:
\begin{align}
\label{eq:full-loc-hypers}
A_i{}^\alpha =&\, \mathring{A}_i{}^\alpha \, , \nonumber \\
H_i{}^\alpha =&\, \mathrm{i}\,\sigma_3{}^\alpha{}_i\,\Bigl(\frac{r - 1}{r+ 1}\Bigr)^{1/2}\,\xi_I (\Sigma^I_+ - \Sigma^I_-) = \mathrm{i}\,\sigma_3{}^\alpha{}_i\,\Bigl(\frac{r - 1}{r + 1}\Bigr)^{-1/2}\,\xi_I (\Sigma^I_+ + \Sigma^I_-) \, .
\end{align}

Before closing this section, we give the explicit BPS gauge field on the localization locus, obtained by integrating the field strengths in \eqref{eq:full-loc-vectors}. In the gauge $W_r^I = 0$, we find
\begin{equation}
\label{eq:full-gauge-field}
W_\mu^I = \Bigl(e^I(r-1) + \sqrt{v_1}\,\sum_{k=2}^\infty (C_k^I+ D^I_{k-1})\bigl(r^{1-k} - 1\bigr) , \, 0 , \, 0 , \, -p^I\cos\theta\Bigr) \, ,
\end{equation}
where we have included the appropriate constants of integration so that the Wilson line $\int W^I_\tau\,\mathrm{d}\tau$ around the thermal circle vanishes at the horizon $r_H=1$. Note that the field configuration \eqref{eq:full-loc-vectors}, \eqref{eq:full-loc-hypers} and \eqref{eq:full-gauge-field} is smooth and regular everywhere provided we keep the sums $\sum_k(C_k^I \pm D_k^I)$ fixed and finite at the horizon.

\section{Supergravity action on the localization locus}
\label{sec:loc-action}

In this section, we evaluate the supergravity action on the localization locus where the off-shell fluctuations of the matter fields are controlled by the functions $\Sigma_\pm^I$ according to the analysis of the previous section. This constitutes step 4 in the program outlined in the introduction. Throughout, we will consider a two-derivative theory controlled by the prepotentials $\mathcal{F}^\pm(X_\pm)$, which we recall are homogeneous functions of degree two in the real scalars of the vector multiplets.

We start with the Lagrangian for vector multiplets given in App. \ref{app:actions}. We evaluate it on the field configuration where the Weyl multiplet fields take their attractor half-BPS values presented in section \ref{sec:on-shell} and the vector multiplet fields are given by \eqref{eq:full-loc-vectors}. We first make a rotation in field space by sending\footnote{Observe that this prescription is identical to the one used in \cite{Benini:2015noa} to treat the scalar auxiliary field of the vector multiplet in the dual CFT$_3$ when using localization to compute the topologically twisted index.}
\begin{equation}
\label{eq:Y-eucl}
Y^{12\,I} \rightarrow \mathrm{i}\,Y^{12\,I}_{(\mathrm{cont.})} \, ,
\end{equation}
where we do not impose a reality condition on the field $Y^{12\,I}_{(\mathrm{cont.})}$. On the localization locus however, it is real and given by 
\begin{equation}
Y^{12\,I}_{(\mathrm{cont.})}(r) = \frac{p^I}{v_2} - \frac{1}{\sqrt{v_1}}\bigl(r\,\partial_r\Sigma^I_+ + \partial_r\Sigma_-^I\bigr)  \, .
\end{equation}

After implementing \eqref{eq:Y-eucl} and some manipulations (involving in particular the homogeneity of the prepotentials), we find that the vector multiplet Lagrangian on the localizing manifold can be put in a simple form, 
\begin{align}
  \label{eq:vec-Lag-1}
e^{-1}\mathcal{L}_\mathrm{V} = \frac{1}{2v_2\sqrt{v_1}}\,p^I\Bigl[&\,N_{IJ}(\partial_r\Sigma^J_- + r\partial_r\Sigma^J_+ + \Sigma^J_+) + (\mathcal{F}_{IJ}^+ - \mathcal{F}_{IJ}^-)(\partial_r\Sigma_+^J + r\partial_r\Sigma^J_- + \Sigma_-^J) \nonumber \\
&\; + 2\,\mathcal{F}_{IJ}^+\,\mathring{X}^J_+ + 2\,\mathcal{F}_{IJ}^-\,\mathring{X}^J_-\Bigr] \, .
\end{align}
where we also made use of the attractor equation \eqref{eq:elec-attract}. Note that in this expression, the matrix $N_{IJ}$ and the prepotentials depend on the fluctuations $\Sigma_\pm^I$.

From the expression \eqref{eq:vec-Lag-1} one can show that the vector Lagrangian can be conveniently written as a total derivative:
\begin{equation}
e^{-1}\mathcal{L}_\mathrm{V} = \frac{1}{v_2\sqrt{v_1}}\,p^I\,\partial_r\bigl[r(\mathcal{F}_I^+ + \mathcal{F}_I^-) + \mathcal{F}_I^+ - \mathcal{F}_I^-\bigr] \, .
\end{equation}
In turn, the action reads
\begin{align}
S_\mathrm{V} =&\, 2\pi\sqrt{v_1} \int_{S^2} \mathrm{d}\theta\,\mathrm{d}\varphi\,\sin\theta \int_1^{r_0}\,\mathrm{d}r\,p^I\partial_r\bigl[r(\mathcal{F}_I^+(X_+) + \mathcal{F}_I^-(X_-)) + \mathcal{F}_I^+(X_+) - \mathcal{F}_I^-(X_-)\bigr] \nonumber \\
=&\, 2\pi\sqrt{v_1}\,p^I \int_{S^2} \mathrm{d}\theta\,\mathrm{d}\varphi\,\sin\theta\Bigl[r_0\bigl(\mathcal{F}_I^+(X_+(r_0)) + \mathcal{F}_I^-(X_-(r_0))\bigr) \\
&\qquad \qquad \qquad \qquad \qquad \; + \mathcal{F}_I^+(X_+(r_0)) - \mathcal{F}_I^-(X_-(r_0)) - 2\,\mathcal{F}_I^+(X_+(r=1))\Bigr] \, , \nonumber 
\end{align}
where we have introduced a radial cut-off $r_0$ to regularize the infinite volume of ${\cal H}_2$. Taking the limit $r_0 \rightarrow \infty$ and Taylor-expanding the prepotentials, we obtain
\begin{align}
&S_\mathrm{V} = 8\pi^2\sqrt{v_1}\,r_0\,p^I\,(\mathring{\mathcal{F}}_I^+ + \mathring{\mathcal{F}}_I^-) \\
&+ 2\pi\sqrt{v_1}\,p^I \int_{S^2} \mathrm{d}\theta\,\mathrm{d}\varphi\,\sin\theta\Bigl[\tfrac12\mathring{N}_{IJ}D_1^J + \tfrac12(\mathring{\mathcal{F}}_{IJ}^+ - \mathring{\mathcal{F}}_{IJ}^-)C_1^J +  \mathring{\mathcal{F}}_I^+ - \mathring{\mathcal{F}}_I^- - 2\,\mathcal{F}_I^+(X_+(r=1))\Bigr] \, , \nonumber
\end{align}
up to terms that vanish in the $r_0 \rightarrow \infty$ limit. The relation between electric and magnetic charges \eqref{eq:q-p-attract}, the attractor equations \eqref{eq:elec-attract}, \eqref{eq:mag-attract} and the homogeneity of the prepotentials can now be used to write the vector multiplet action in the following form
\begin{align}
&S_\mathrm{V} = 8\pi^2\sqrt{v_1}\,r_0\,p^I\,(\mathring{\mathcal{F}}_I^+ + \mathring{\mathcal{F}}_I^-) \\
&- 2\pi\sqrt{v_1}\int_{S^2} \mathrm{d}\theta\,\mathrm{d}\varphi\,\sin\theta\Bigl[p^I\mathcal{F}_I^+\bigl(2X_+(r=1)\bigr) + q_I\bigl(\mathring{X}^I_+ + \mathring{X}^I_- + C_1^I\bigr) - v_2\,\chi_{\mathrm{H}}\,\xi_I D_1^I\Bigr] + \mathcal{O}(1/r_0) \, . \nonumber
\end{align} 

Next, we turn to the Wilson line contribution to the action. It can be directly evaluated from the value of the gauge field \eqref{eq:full-gauge-field} at the boundary $r=r_0$. Because of the BPS conditions \eqref{eq:vec-BPS-ang-gen}, the $\tau$ component of the gauge field only depends on the radial coordinate. Therefore, we obtain
\begin{equation}
\label{eq:S_W}
S_\mathrm{W} = 4\pi\,q_I \int_0^{2\pi} W^I_\tau(r_0)\,\mathrm{d}\tau = 8\pi^2\,q_I e^I\,(r_0-1) - 8\pi^2\sqrt{v_1}\,q_I \sum_{k=2}^\infty (C_k^I+ D^I_{k-1}) + \mathcal{O}(1/r_0) \, .
\end{equation}

Finally, we discuss the compensating hypermultiplet contribution. Even though this multiplet is introduced to ensure that the conformal theory is gauge-equivalent to Poincar\'{e} supergravity, the gauging introduces a coupling between the hypermultiplet fields and the vector multiplets, which in turn gives a non-trivial dependence on the off-shell fluctuations to the total action of the theory evaluated on the localization locus, as we now show.

Since we treat the compensating hypermultiplet off-shell with respect to the localizing supercharge, its Lagrangian \eqref{eq:hyper-action} must be supplemented by the standard quadratic term for the auxiliary fields $H_i{}^\alpha$,
\begin{equation}
e^{-1}\mathcal{L}_{\mathrm{aux.}} = \frac12\,\chi_\mathrm{H}\,H_i{}^\alpha H^i{}_\alpha \, .
\end{equation}
After implementing the rotation \eqref{eq:Y-eucl}, we can evaluate $\mathcal{L}_\mathrm{H} + \mathcal{L}_{\mathrm{aux.}}$ on the localization locus using \eqref{eq:full-loc-vectors} and \eqref{eq:full-loc-hypers}. All the quadratic terms in the fluctuations cancel, and we are left simply with
\begin{equation}
\label{eq:hyper-lag}
e^{-1}(\mathcal{L}_\mathrm{H} + \mathcal{L}_{\mathrm{aux.}}) = -\frac{1}{\sqrt{v_1}}\,\chi_\mathrm{H}\,\xi_I\bigl(\Sigma_+^I + r\partial_r\Sigma_+^I + \partial_r\Sigma_-^I\bigr) \, .
\end{equation}
This can be integrated to give the hypermultiplet action, including the contribution from the auxiliary fields,
\begin{equation}
S_{\mathrm{H}} = -2\pi\,v_2\sqrt{v_1}\,\chi_\mathrm{H}\int_{S^2} \mathrm{d}\theta\,\mathrm{d}\varphi\,\sin\theta\Bigl[\xi_I D_1^I - \xi_I C_1^I - \sum_{k=1}^\infty \xi_I(C^I_{k+1} + D^I_k)\Bigr] + \mathcal{O}(1/r_0) \, ,
\end{equation}
in the limit where the radial cut-off $r_0 \rightarrow \infty$. Now, recalling the constraint \eqref{eq:full-loc-Ck-constraint}, this can be further simplified and we are left with
\begin{equation}
S_{\mathrm{H}} = -2\pi\,v_2\sqrt{v_1}\,\chi_\mathrm{H}\int_{S^2} \mathrm{d}\theta\,\mathrm{d}\varphi\,\sin\theta\,\xi_I D_1^I + \mathcal{O}(1/r_0) \, .
\end{equation}

We must now combine all the above contributions. For this, we note that the BPS equations \eqref{eq:vec-BPS-ang-gen} allow us to write the prefactor in the second term of $S_{\mathrm{W}}$ \eqref{eq:S_W} as a volume integral on $S^2$. Doing so, the bulk action of the Euclidean theory describing $n_V + 1$ vector multiplets and the compensating hypermultiplet coupled to the Weyl background $S_{\mathrm{bulk}} := S_{\rm V} + S_{\rm H}$, and supplemented by the Wilson line $S_{\rm W}$, evaluates on the localization locus to
\begin{align}
\label{eq:S_tot-ang}
&S_\mathrm{bulk} + S_{\rm W} = \\
&\; 8\pi^2\sqrt{v_1}\,r_0\,\Bigl[p^I(\mathring{\mathcal{F}}_I^+ + \mathring{\mathcal{F}}_I^-) + v_1^{-1/2}\,q_I e^I \Bigr] - 2\pi\sqrt{v_1}\int_{S^2} \mathrm{d}\theta\,\mathrm{d}\varphi\,\sin\theta\,\Bigl[p^I\mathcal{F}_I^+(\phi_+)+ q_I\phi_+^I \Bigr] \, , \nonumber
\end{align}
up to terms vanishing when the cut-off $r_0$ goes to infinity. Above, we have introduced the following real variable, which depends on the fluctuation parameters $C_k^I(\theta,\varphi)$ and $D_k^I(\theta,\varphi)$,
\begin{equation}
\label{eq:phi-var}
\phi_+^I := 2X_+(r=1) = 2\mathring{X}^I_+ + \sum_{k=1}^\infty(C_k^I + D_k^I) \, .
\end{equation}
Observe that, owing to \eqref{eq:vec-BPS-ang-gen}, this variable is in fact constant on $S^2$. We can therefore evaluate the angular integration straightforwardly in \eqref{eq:S_tot-ang} to obtain the final form of the supergravity action on the localization locus:
\begin{equation}
\label{eq:S_tot}
S_\mathrm{bulk} + S_{\rm W} = 8\pi^2\sqrt{v_1}\,r_0\,\Bigl[p^I(\mathring{\mathcal{F}}_I^+ + \mathring{\mathcal{F}}_I^-) + v_1^{-1/2}\,q_I e^I \Bigr] - 8\pi^2\sqrt{v_1}\,\Bigl[p^I\mathcal{F}_I^+(\phi_+)+ q_I\phi_+^I \Bigr] \, .
\end{equation}
As expected, the total action \eqref{eq:S_tot} contains a term divergent in the cut-off $r_0$ due to the infinite volume of $\mathcal{H}_2$. We therefore need to add boundary counter-terms in order to regularize this divergence.

\subsection{Boundary counter-terms and supersymmetry}
\label{subsec:holo-renorm}

A first-principle derivation of the appropriate boundary counter-terms requires a careful analysis of the theory on $\mathcal{H}_2$ in order to ensure that the variational problem for the QEF \eqref{eq:macroscopic-degeneracy} is well-defined (see \cite{Castro:2008ms,Grumiller:2017qao}) and that the renormalization procedure preserves supersymmetry. Indeed, besides the usual Gibbons-Hawking-York term, it is well-known that supersymmetry may require the addition of finite terms which modify the renormalized bulk action (see e.g. \cite{Freedman:2013ryh} for a discussion in the context of four-dimensional theories with three-dimensional boundaries). In this paper we will instead employ a ``minimal'' subtraction scheme and show that the naive boundary counter-terms we add can be arranged in a manner which makes supersymmetry of the resulting renormalized action manifest.

To cancel the divergences in \eqref{eq:S_tot}, we add the following counter-terms at the boundary:
\begin{equation}
S_{\rm bct} = S_{\rm bct}^1 + S_{\rm bct}^2 \, ,
\end{equation}
with
\begin{align}
S_{\rm bct}^1 :=&\, -8\pi^2\,\sqrt{v_1}\,r_0\,p^I(\mathring{\mathcal{F}}_I^+ + \mathring{\mathcal{F}}_I^-) \, , \\
S_{\rm bct}^2 :=&\, -8\pi^2\,r_0\,q_I\,e^I \, .
\end{align}
We then choose the renormalized action
\begin{equation}
\label{eq:S_ren}
S_{\rm ren} := \big(S_{\rm bulk} + S_{\rm bct}^1\big) + \big(S_{\rm W}+S_{\rm bct}^2\big) \, ,
\end{equation}
Upon using the attractor equation \eqref{eq:elec-attract}, the second term above can be interpreted as the supersymmetric Wilson line of the gauge theory at the boundary in the limit where $r_0\to\infty$,
\begin{equation}
S_{\rm W} + S_{\rm bct}^2 = 4\pi\,q_I \,\int_0^{2\pi} (W_\tau^I(r_0) - e^1_\tau(\mathring{X}_+^I - \mathring{X}_-^I)\bigr)\,\mathrm{d}\tau \, ,
\end{equation}
where $e_\tau^1=\sqrt{v_1(r_0^2-1)}$ is the component of the vielbein $e_{\mu}^a$ evaluated at the boundary. It is easy to realize that the above expression coincides with the supersymmetric invariant Wilson line at the boundary \cite{Maldacena:1998im}, and that it is manifestly finite.\\
Analogously, it is possible to show that the supersymmetric variation of $S_{\rm bulk}+S_{\rm bct}^1$ vanishes, since the bulk action varies into a total derivative which, when evaluated in the limit where $r_0\to\infty$, exactly cancels the total derivative term obtained by varying the boundary counter-term $S_{\rm bct}^1$ \cite{Dabholkar:2010uh}. We refer the reader to Appendix D of \cite{Dabholkar:2010uh} for a proof of this statement and further details.\footnote{Note that the analysis \cite{Dabholkar:2010uh} in ungauged supergravity only relies on the vector and Weyl multiplet transformation rules, which are identical to the ones of the gauged supergravity theory we consider here. We thus expect their results to apply straightforwardly to our case.}\\
We stress again that the minimal boundary counter-terms we choose, though consistent with supersymmetry at the boundary, a priori carry no information about potential finite terms in the action. However, as we will show below, the final result we obtain from the above renormalization scheme is consistent without any finite terms. If such terms do exist, their form will therefore be highly constrained.

The conclusion of this subsection is that we will use the renormalized action \eqref{eq:S_ren},
\begin{equation}
\label{eq:S_final}
S_{\rm ren}[\phi_+] = - 8\pi^2\sqrt{v_1}\,\Bigl[p^I\mathcal{F}_I^+(\phi_+) + q_I\phi_+^I \Bigr] \, , 
\end{equation}
for the evaluation of the QEF, to which we turn to now.

\subsection{Finite result and saddle-point evaluation}
\label{subsec:finite}

We are now ready to write the formula for the localized QEF and bring to a conclusion step 4 of the localization program outlined in the introduction. The localization locus of step 3 is parameterized by $2n_V + 2$ infinite sets of functions $\bigl(C_k^I(\theta,\varphi),D_k^I(\theta,\varphi)\bigr)$ with $k\geq1$ and $I=0\ldots n_V$. They are constrained according to \eqref{eq:full-loc-C1-constraint} and \eqref{eq:full-loc-Ck-constraint}. As a result, we have that the QEF \eqref{eq:macroscopic-degeneracy} is equal to a constrained functional integral
\begin{equation}
d_{\rm macro}(p^I,q_I) = \int' \,\Bigl(\prod_{I=0}^{n_V}\,\prod_{k=1}^\infty\,\mathcal{D}C_k^I\,\mathcal{D}D_k^I\Bigr)\;e^{S_{\rm ren}[\phi_+]}\,Z_{\rm ind}(p^I,q_I;C_k,D_k) \, ,
\end{equation}
where the prime on the integral means that it must be computed enforcing the constraints \eqref{eq:full-loc-C1-constraint} and \eqref{eq:full-loc-Ck-constraint}, $\phi_+$ is defined in \eqref{eq:phi-var} and $Z_{\rm ind}$ denotes the contribution from the one-loop determinant of quadratic fluctuations around the localization locus and an eventual measure factor\footnote{This measure factor is known to be non-trivial in the case of asymptotically flat black holes (see e.g. \cite{Dabholkar:2011ec,Murthy:2015zzy}), as well as in the case of the partition function for AdS$_4$ spaces \cite{Dabholkar:2014wpa} and the topological black hole studied in \cite{Nian:2017hac}.}.

Observe that the renormalized action $S_{\rm ren}$ entering the integrand only depends on a particular linear combination of the integration variables $C_k^I$ and $D_k^I$, namely $\phi_+^I$, and that this variable is in fact independent of the angular coordinates owing to the BPS equations. Let us denote the directions orthogonal to the $\phi^I_+$ line in the $(C_k^I,D_k^I)$-space by $\phi_0^I(\theta,\varphi)$. Note that while $\phi_+^I$ is constant on $S^2$ thanks to \eqref{eq:vec-BPS-ang-gen}, $\phi_0^I(\theta,\varphi)$ is still a function of the angular coordinates in general. We can then write the exact\footnote{Recall from the comment at the end of section \ref{subsec:vector-fluct} that we only consider the contributions from non-singular field configurations for the localizing manifold in this paper.} macroscopic degeneracy as
\begin{align}
\label{eq:dmacro-loc-diverge}
&d_{\rm macro}(p^I,q_I) = \\
&\quad \int_{-\infty}^{+\infty} \,\Bigl(\prod_{I=0}^{n_V}\,\mathrm{d}\phi_+^I\Bigr)\;\delta\Bigl(\xi_I\phi^I_+ - \frac{1}{2\sqrt{v_1}}\Bigr)\,e^{S_{\rm ren}[\phi_+]} \int'\,\Bigl(\prod_{I=0}^{n_V}\,\mathcal{D}\phi_0^I\Bigr)\,Z_{\rm ind}(p^I,q_I,\phi_+;\phi_0) \, , \nonumber
\end{align}
where, in the first integral, we have enforced the constraint on $\phi_+^I$ stemming from \eqref{eq:full-loc-C1-constraint} and \eqref{eq:full-loc-Ck-constraint} using a delta function. The last integral is infinite-dimensional, and as such one may be worried that it renders \eqref{eq:dmacro-loc-diverge} infinite or ill-defined. To properly address this issue would require us to complete step 5 of the localization program, in order to obtain explicit expressions for the one-loop determinants and the measure as functions of the $\phi_+^I$ and $\phi_0^I$ parameters. However we can already sketch an argument (based on supersymmetry) which shows that this integral will likely be regular, by examining the details of the localization calculation in the CFT$_3$ dual to our black hole asymptotics. It was shown in \cite{Benini:2015noa} that, for the correct localization of the topologically twisted index, one must take into account the interplay between bosonic and fermionic zero-modes on the localization locus. In our situation, we can interpret the $\phi_0^I(\theta,\varphi)$ as bosonic zero-modes on the locus, since our renormalized action does not depend on them. We should therefore ask what is the effect of including their fermionic superpartners under the localizing supercharge $Q_{\rm loc}$, which we denote by $\Omega^I_0(\theta,\varphi)$ since they will arise from the gaugini in the vector multiplets. The effect of including these fermionic zero-modes is to modify the second integral in \eqref{eq:dmacro-loc-diverge} to
\begin{equation}
\label{eq:Z-ind-reg}
Z_{\mathrm{ind}}^{\rm reg}(p^I,q_I,\phi_+) :=  \int'\,\Bigl(\prod_{I=0}^{n_V}\,\mathcal{D}\phi_0^I\,\mathcal{D}\Omega_0^I\Bigr)\,Z_{\rm ind}(p^I,q_I,\phi_+;\phi_0,\Omega_0) \, .
\end{equation}
Due to the supersymmetric pairing under $Q_{\rm loc}$ of the bosonic and fermionic zero-modes, we expect that the (constrained) integration in \eqref{eq:Z-ind-reg} will reduce to a regular function of $\phi^I_+$ and of the black hole charges \cite{Benini:2015noa}.\footnote{More explicitly, we expect supersymmetry to relate the derivative of $Z_{\rm ind}$ with respect to the fermionic zero-modes $\Omega_0^I$ to its derivative with respect to $\phi^I_0$ since the modes are paired. The integral over the $\phi_0$ modes is then a total derivative, and we assume that it is a regular function of $\phi_+$ and the charges.} Evidently, for the reasons already mentioned, we cannot rigorously prove the cancellation of the divergences resulting from the integration over the zero-modes at this stage. This will be examined in future work. Nevertheless, let us remark that our situation in gauged supergravity parallels the situation in the CFT$_3$ of \cite{Benini:2015noa}: due to the gauging, we effected a twist in the S$^2$ factor of the near-horizon geometry which resulted in constant Killing spinors along the sphere (see App. \ref{app:CKS}). As such, the gauging and the half-BPS character of the near-horizon configuration introduce an extra degeneracy as compared to the case of the ungauged, full-BPS near-horizon configuration studied in \cite{Dabholkar:2010uh}. This ultimately manifests itself in the fact that the infinite tower of parameters $C_k^I$ and $D_k^I$ with $k\geq 1$ is not truncated in our situation.\\

With this motivation in mind, we write the finite result for the localized QEF as\footnote{To obtain this form we have rescaled the variables $\phi_+^I \rightarrow \tfrac{\phi_+^I}{4\pi\sqrt{v_1}}$ in \eqref{eq:dmacro-loc-diverge} and absorbed the resulting constant prefactor in $Z^{\rm reg}_{\rm ind}$. With this change, the newly defined integration variables in \eqref{eq:dmacro-loc-reg} have zero mass dimension.}
\begin{equation}
\label{eq:dmacro-loc-reg}
d_{\rm macro}(p^I,q_I) = \int_{-\infty}^{+\infty} \,\Bigl(\prod_{I=0}^{n_V}\,\frac{\mathrm{d}\phi_+^I}{2\pi}\Bigr)\delta\bigl(\xi_I\phi^I_+ - 2\pi\bigr)\exp\Bigl[-2\pi\bigl(p^I\mathcal{F}_I^+(\phi_+) + q_I\phi_+^I\bigr)\Bigr]Z^{\rm reg}_{\rm ind}(\phi_+) \, ,
\end{equation}
The expression \eqref{eq:dmacro-loc-reg} is general since it applies to any black hole solution of two-derivative $\mathcal{N}=2$ gauged superconformal gravity governed by a prepotential $\mathcal{F}^+(X_+)$ and having AdS$_2 \times $S$^2$ near-horizon geometry.

The saddle-point evaluation of \eqref{eq:dmacro-loc-reg} proceeds by defining
\begin{equation}
\mathcal{Z}(\phi_+) := -2\pi\,\bigl(p^I\mathcal{F}_I^+(\phi_+) + q_I \phi^I_+\bigr) \, , \qquad \mathcal{L}(\phi_+) := \frac{1}{2\pi}\,\xi_I\phi^I_+ \, .
\end{equation} 
In terms of these quantities, the constraint \eqref{eq:constraint-hypers} tells us that $\mathcal{L}=1$ and the saddle-point equations coincide with the standard attractor equations\footnote{Note that we have so far only taken electric gauging parameters $\xi_I$ and set their magnetic counterparts $\xi^I = 0$. This is the reason why we do not recover the full symplectic covariant form of the attractor mechanism \cite{Dall'Agata:2010gj}, where
\begin{equation}
	\mathcal{L}(\phi_+) := \bigl(\xi_I\phi^I_+ + \xi^I \mathcal{F}_I^+(\phi_+)\bigr)/2\pi \, . \nonumber
\end{equation}
However, we expect to find the symplectically covariant version straightforwardly by repeating our steps starting from the dyonically gauged off-shell supergravity formalism of \cite{deWit:2011gk}.} \cite{Cacciatori:2009iz,Dall'Agata:2010gj}
\begin{equation}
\frac{\partial}{\partial \phi_+^I}\,\frac{\mathcal{Z}}{\mathcal{L}} = 0 \, .
\end{equation}
This shows that the saddle-point is located at the on-shell value $\mathring{\phi}^I_+ = 8\pi\sqrt{v_1}\,\mathring{X}^I_+$. Furthermore, the value of $\mathcal{Z}/\mathcal{L}$ at this extremum reproduces the Bekenstein-Hawking entropy, 
\begin{equation}
\frac{\mathring{\mathcal{Z}}}{\mathring{\mathcal{L}}} = -4\pi^2\,v_2\,\chi_\mathrm{H} = \frac{4\pi\,v_2}{4\,\mathcal{G}_{\rm N}} = \frac{A_\mathrm{H}}{4\,\mathcal{G}_{\rm N}}\, , 
\end{equation}
where we have used that $\chi_\mathrm{H} = -(4\pi\,\mathcal{G}_{\rm N})^{-1}$ with $\mathcal{G}_{\rm N}$ the Newton constant (see the next subsection), and $A_{\rm H}$ is the area of the horizon (cf. \eqref{eq:1/2-BPS-metric}). The fact that we obtain the usual attractor mechanism and leading entropy in the saddle-point approximation from the exponent of \eqref{eq:dmacro-loc-reg} implies that the factor $Z_{\rm ind}^{\rm reg}$ can contribute at most $\log(A_{\rm H})$ terms in the large horizon limit.

\subsection{Lorentzian variables}
\label{subsec:FT}

In the next section, we will compare the macroscopic degeneracy above to the corresponding microscopic degeneracy obtained in the dual field theory for various models. Therefore, we dedicate the rest of this section to massaging \eqref{eq:dmacro-loc-reg} by making various field and charge redefinitions to bring it to a Lorentzian form which allows for a more immediate comparison with its holographic counterpart. 

Recall that the bulk supergravity Lagrangian used to derive \eqref{eq:dmacro-loc-reg} contains the following terms (see again App. \ref{app:actions})
\begin{equation}
e^{-1}\mathcal{L}_{\rm bulk} = \frac16R(\chi_{\rm H} - \chi_{\rm V}) + D(\chi_{\rm V} + \tfrac12\chi_{\rm H}) + N_{IJ}\partial_\mu X_+^I\partial^\mu X_-^J - 4\chi_{\rm H}\,g^2(\xi_I X_+^I)(\xi_J X^J_-) + \ldots \, ,
\end{equation}
where we have reinstated the coupling constant $g$ for the gauging (see below \eqref{eq:hyper-4D}). Note that since $g$ has mass dimension 1, $\xi_I$ in the expression above has mass dimension $-1$. The equation of motion for the $D$ auxiliary field sets $\chi_{\rm H} = -2\chi_{\rm V}$, and we obtain the canonically normalized Einstein-Hilbert Lagrangian (reinstating Newton's constant $\mathcal{G}_{\rm N}$) by fixing the dilatation gauge and choosing $\chi_{\rm V} = 1/\kappa^2$ where $\kappa^2 := 8\pi\mathcal{G}_{\rm N}$,
\begin{equation}
e^{-1}\mathcal{L}_{\rm bulk} = -\frac{1}{2\kappa^2}R + N_{IJ}\partial_\mu X_+^I\partial^\mu X_-^J - 8\,\frac{g^2}{\kappa^2}(\xi_I X_+^I)(\xi_J X^J_-) + \ldots \, .
\end{equation}
We can switch to a more standard holographic convention by redefining the scalar fields and FI terms to extract an overall factor of $1/2\kappa^2$ in the bulk Lagrangian, 
\begin{equation}
e^{-1}\mathcal{L}_{\rm bulk} = \frac{1}{2\kappa^2}\Bigl[-R + N_{IJ}\partial_\mu \widetilde{X}_+^I\partial^\mu \widetilde{X}_-^J - 8\,g^2(\widetilde{\xi}_I \widetilde{X}_+^I)(\widetilde{\xi}_J \widetilde{X}^J_-)\Bigr] + \ldots \, ,
\end{equation}
where
\begin{equation}
\label{eq:rescale}
\widetilde{X}^I_\pm := \sqrt{2}\,\kappa\,X_\pm^I \, , \quad \widetilde{\xi}_I := \kappa^{-1}\,\xi_I \, ,
\end{equation}
are dimensionless scalar fields and FI parameters, which are most suited for holography. 

Aside from the fields and gauging parameters, recall that the charges $(p^I,q_I)$ in the supergravity theory used to derive \eqref{eq:dmacro-loc-reg} are normalized as follows (see \eqref{eq:def-dual}),
\begin{equation}
\int_{S^2} \mathring{F}^I = 4\pi\,p^I \, , \quad \int_{S^2} \mathring{G}_I = 4\pi\,q_I \, ,
\end{equation}
and the resulting lattice of electro-magnetic charges is
\begin{equation}
\label{eq:EM-lattice}
2\,g\,\xi_I\,p^I \in \mathbb{Z} \, , \quad \frac{4\pi}{2\,g\,\xi_I}\,q_I \in \mathbb{Z} \, , 
\end{equation}
where no sum over $I$ is implied. If we now define the rescaled charges,
\begin{equation}
\label{eq:FT-charges}
\widetilde{p}^I := \kappa\,p^I \, , \quad \widetilde{q}_I := \kappa\,q_I \, ,
\end{equation}
they satisfy the quantization conditions 
\begin{equation}
\label{eq:EM-lattice-2}
2\,\widetilde{\xi}_I\,g\,\widetilde{p}^I \in \mathbb{Z} \, , \quad (2\,\widetilde{\xi}_I)^{-1}\frac{\widetilde{q}_I}{2\,g\,\mathcal{G}_{\rm N}} \in \mathbb{Z} \, ,
\end{equation}
in terms of the dimensionless FI parameters \eqref{eq:rescale}. 

Changing variables in \eqref{eq:dmacro-loc-reg} from $\phi^I_+$ to $2(g\kappa)^{-1}\phi^I_+$, using the dimensionless FI parameters and the charges $(\widetilde{p}^I,\widetilde{q}_I)$, we obtain
\begin{equation}
\label{eq:dmacro-FT}
d_{\rm macro}(\widetilde{p}^I,\widetilde{q}_I) = \int_{-\infty}^{+\infty} \Bigl(\prod_{I=0}^{n_V}\,\frac{\mathrm{d}\phi_+^I}{2\pi}\Bigr)\delta\bigl(2\widetilde{\xi}_I\phi^I_+ - 2\pi\bigr)\exp\Bigl[-\frac{1}{2\,g\,\mathcal{G}_{\rm N}}\Bigl(\widetilde{p}^I\mathcal{F}_I^+(\phi_+) + \widetilde{q}_I\phi_+^I\Bigr)\Bigr]Z^{\rm reg}_{\rm ind}(\phi_+) \, ,
\end{equation}
where we dropped a dimensionless constant $(2/g\kappa)^{n_V+1}$ coming from the change of variables. In the next section, we will see that the charges $(g \widetilde{p}^I, \widetilde{q}_I/2g\mathcal{G}_{\rm N})$ should be identified with the charges in the dual field theory. Note that in \eqref{eq:dmacro-FT} we have included the coupling constant $g$ and Newton's constant $\mathcal{G}_{\rm N}$ which have mass dimensions $1$ and $-2$, respectively. The integration variables $\phi_I^+$ and the FI parameters $\widetilde{\xi}_I$ have mass dimensions zero while the charges $(\widetilde{p}^I,\widetilde{q}_I)$ have mass dimension $-1$.

In this formulation, we come back to the Wilson line which is inserted at the boundary of ${\cal H}_2$ in the Euclidean QEF \eqref{eq:macroscopic-degeneracy}. We see using \eqref{eq:FT-charges} that, in order to have the correct normalization for the gauge-invariant Wilson line in Lorentzian signature, we must send $\widetilde{q}_I \rightarrow \mathrm{i}\,\widetilde{q}_I$. This shows that the correct quantity which computes the entropy of the physical black hole, and that we will compare to its holographic counterpart in the next section, is an analytic continuation 
\begin{equation}
\label{eq:dmacro-Lor}
d_{\rm macro}(\widetilde{p}^I, \mathrm{i}\,\widetilde{q}_I) \, . 
\end{equation}
We can interpret this analytic continuation as an artifact of using a Euclidean expectation value to compute the macroscopic degeneracy of the Lorentzian black hole solution. Along these lines, the natural expectation is that the fluctuation parameters $C^I_k$ are sent to $\mathrm{i}\,C^I_k$, while $D^I_k$ are not analytically continued. Therefore the Lorentzian macroscopic degeneracy is given by $n_V$ complex integrations over the complex plane.

\section{Holographic examples}
\label{sec:examples}

It was argued in \cite{Benini:2015eyy,Benini:2016rke} that, in cases when the holographic correspondence is available, one should consider the topologically twisted index \cite{Benini:2015noa} of the field theory dual in order to describe microscopically the black hole degrees of freedom. The twisted index is the partition function of a topologically twisted theory\footnote{Just as we can consider any $\Sigma_g$ for the horizon of the black hole, we can also consider a topologically twisted holographic dual on a general $\Sigma_g\times$S$^1$ \cite{Benini:2016hjo,Closset:2016arn}. The final result is again a straightforward generalization of the spherical case, and therefore in this section we again specialize to spherical topology.} on S$^2 \times$S$^1$ and depends on a set of magnetic fluxes $\mathfrak{p}_a$ (holographically corresponding to the properly normalized black hole magnetic charges) and chemical potentials $\Delta_a$ for the global symmetries $J_a$ of the theory (holographically corresponding to the gauge symmetries in supergravity, a number of $U(1)$'s). The twisted index can be interpreted as the Witten index which counts ground states in the dimensionally reduced quantum mechanics holographically describing the degrees of freedom near the black hole horizon,
\begin{equation}  
	Z(\mathfrak{p}_a, \Delta_a) = \text{Tr} (-1)^F\ e^{-\beta H}\ e^{i \Delta_a J_a}\ ,
\end{equation}
where supersymmetry further leads to the restrictions
\begin{equation}
	\sum_a \mathfrak{p}_a = 2\ , \qquad \sum_a \Delta_a = 2 \pi\ .
\end{equation}
Finally, to evaluate the number of supersymmetric ground states $d_{\rm micro}$ as a function of the electric and magnetic charges one needs to perform a Fourier transform with respect to the chemical potentials,
\begin{equation}\label{eq:micro_entropy}
	d_{\rm micro} (\mathfrak{p}_a,\mathfrak{q}_a) = \int_0^{2\pi} \left(\prod_a \frac{\mathrm{d}\Delta_a}{2 \pi} \right) \delta\Bigr(\sum_a \Delta_a - 2\pi\Bigl)\ Z(\mathfrak{p}_a, \Delta_a)\ e^{-i \sum_a \Delta_a \mathfrak{q}_a}\ ,
\end{equation}
which formally gives the {\it exact} holographic expression for the microscopic entropy of the black hole. This is therefore the expression that we would ultimately like to match with the exact evaluation of the quantum entropy function. While our result for the QEF \eqref{eq:dmacro-FT} bears some important similarities with the microscopic degeneracy \eqref{eq:micro_entropy}, we are not yet in a position to successfully match the expressions at all orders. Here it is important to note that, unlike the QEF which can be evaluated for a general prepotential, the evaluation of the twisted index $Z(\mathfrak{p}_a, \Delta_a)$ proceeds very differently for the different holographic examples, which we briefly summarize next. In any case, due to the fact that we have not yet completed the final step 5 in our localization program which consists in evaluating $Z^{\rm reg}_{\rm ind}$, it is clear that any exact holographic match would be too premature and speculative at this stage. We nevertheless comment on how much progress one can hope to make in each of the explicit cases below\footnote{Note that in all examples below we deal with models embeddable in different versions of maximal ${\cal N}=8$ supergravity with 32 supercharges. This number of supersymmetries forces all higher-derivative invariants to vanish and therefore it is natural to expect that these models do not receive higher-derivative corrections from string theory, justifying our comparison with the QEF computed in a two-derivative theory.}.

\subsection{Cacciatori-Klemm black holes}
\label{subsec:CKexample}

The Cacciatori-Klemm black holes \cite{Cacciatori:2009iz} are the supersymmetric solutions of the electrically gauged $STU$ model, specified by the prepotentials (in Euclidean supergravity conventions)
\begin{equation}
\label{eq:prepot-CK}
	\mathcal{F}^{\pm}(X_\pm)_{\rm CK} = \sqrt{X^0_{\pm} X^1_{\pm} X^2_{\pm} X^3_{\pm}}\, ,
\end{equation}
and dimensionless FI parameters
\begin{equation}
\label{eq:FI-CK}
	\widetilde{\xi}_0 = \widetilde{\xi}_1 = \widetilde{\xi}_2 = \widetilde{\xi}_3 = \frac12\, .
\end{equation}
This model arises from a consistent truncation of maximal gauged 4d supergravity and therefore the supersymmetric black holes (with four electric and four magnetic charges) can be uplifted to full eleven-dimensional solutions with the interpretation of M2 branes wrapped on 2-cycles. The holographically dual theory on $N$ coincident M2 branes is a $U(N)\times U(N)$ Chern-Simons theory known as ABJM \cite{Aharony:2008ug}. Its twisted index was evaluated via localization in the large $N$ approximation in \cite{Benini:2015eyy} and, upon extremization, was shown to agree with the leading Bekenstein-Hawking entropy of the Cacciatori-Klemm black holes. The first finite $N$ corrections in the large $N$ limit were more recently considered in \cite{Liu:2017vll}. Therefore at present the known expression for the ABJM twisted index takes the form
\begin{equation}
\label{eq:micro-CK-Z}
 	Z_{\rm ABJM} (\mathfrak{p}_a, \Delta_a) = \exp\Bigl[-\sum_{a=1}^4 \mathfrak{p}_a \frac{\partial {\cal F}_{\rm ABJM}}{\partial \Delta_a}  + N^{1/2} f (\mathfrak{p}_a, \Delta_a) - \frac12 \log N + {\cal O} (N^0)\Bigr] \ ,
\end{equation}
where $f$ is a function that remains as-of-yet undetermined and ${\cal F}_{\rm ABJM}$ is the partition function of ABJM theory on S$^3$,
\begin{equation}
		{\cal F}_{\rm ABJM} (\Delta_a) = \frac{(2 N)^{3/2}}{3} \sqrt{\Delta_1 \Delta_2 \Delta_3 \Delta_4}\, .
\end{equation}
To compare with our result \eqref{eq:dmacro-FT}, we can use the AdS/CFT dictionary for this model which relates Newton's constant to the rank of the gauge group. To leading order in $N$,
\begin{equation}
\frac{1}{2\,g^2\,\mathcal{G}_{\rm N}} = \frac{(2N)^{3/2}}{3} \, .
\end{equation}
Using this, \eqref{eq:prepot-CK} and \eqref{eq:FI-CK}, the localized QEF \eqref{eq:dmacro-FT} for the Cacciatori-Klemm black holes takes the form
\begin{equation}
\label{eq:dmacro-CK}
d_{\rm macro}(\widetilde{p}^I, \mathrm{i}\,\widetilde{q}_I) = \int_{-\infty}^{+\infty}\Bigl(\prod_{I=0}^{3}\,\frac{\mathrm{d}\phi_+^I}{2\pi}\Bigr)\delta\Bigl(\sum_{I=0}^3\phi^I_+ - 2\pi\Bigr)\,Z_{\rm CK}(\widetilde{p}^I,\phi_+)\,e^{-\mathrm{i}\sum_{I=0}^3\tfrac{\widetilde{q}_I}{2g\mathcal{G}_{\rm N}}\phi_+^I} \, ,
\end{equation}
where we defined
\begin{equation}
\label{eq:macro-CK-Z}
Z_{\rm CK}(\widetilde{p}^I,\phi_+) := \exp\Bigl[-\sum_{I=0}^3\,g\,\widetilde{p}^I\,\frac{(2N)^{3/2}}{3}\frac{\partial\sqrt{\phi^0_+\phi^1_+\phi^2_+\phi^3_+}}{\partial \phi^I_+} + \log Z^{\rm reg}_{\rm ind}(\phi_+)\Bigr] \, .
\end{equation}
Comparing \eqref{eq:micro_entropy}, \eqref{eq:micro-CK-Z} with \eqref{eq:dmacro-CK}, \eqref{eq:macro-CK-Z}, we observe a correspondence between macroscopics and microscopics upon identifying the fugacities $\Delta_a$ and the integration variables $\phi^I_+$, and the charges as (note that the r.h.s is properly quantized according to \eqref{eq:EM-lattice-2})
\begin{equation}
\label{eq:charge-match}
(\mathfrak{p}_a,\,\mathfrak{q}_a) = \Bigl(g\,\widetilde{p}^I,\,\frac{\widetilde{q}_I}{2\,g\,\mathcal{G}_{\rm N}}\Bigr) \, .
\end{equation}
At leading order in $N$ the saddle-point approximation gives an exact match, while the exact answer is yet to be determined on each side.

\subsection{Benini-Bobev black strings}
\label{subsec:BBexample}

The class of magnetically charged supersymmetric black strings in AdS$_5$ analyzed holographically in \cite{Benini:2013cda} has a near-horizon geometry AdS$_3 \times$S$^2$. Upon a further dimensional reduction along a periodic coordinate which is part of AdS$_3$ on the horizon, these solutions can be seen as BPS black holes in 4d supergravity (with only up to three of the possible four magnetic charges switched on) with a non-maximally symmetric asymptotic vacuum, as shown in \cite{Hristov:2014eza}. The resulting four-dimensional supergravity model is specified by the prepotentials
\begin{equation}\label{eq:prepot-BB}
	\mathcal{F}^{\pm}(X_\pm)_{\rm BB} = \frac{ X^1_{\pm} X^2_{\pm} X^3_{\pm}}{X^0_{\pm}}\, ,
\end{equation}
and the dimensionless FI terms
\begin{equation}\label{eq:FI-BB}
	\widetilde{\xi}_1 = \widetilde{\xi}_2 = \widetilde{\xi}_3 = \frac12\, ,
\end{equation}
while $\widetilde{\xi}_0$ can be left arbitrary. This model eventually arises from a consistent truncation of maximal gauged 5d supergravity and therefore uplifts to asymptotically AdS$_5 \times$S$^5$ type IIB solutions, thus allowing for an interpretation as wrapped D3 branes on 2-cycles. The holographically dual theory, $\mathrm{SU}(N)$ ${\cal N}=4$ SYM in four dimensions, when put on a Riemann surface, exhibits an RG flow to a two-dimensional superconformal field theory with a central charge that was calculated in \cite{Benini:2013cda} in the large $N$ limit. The entropy in the Cardy limit was then matched successfully with the leading Bekenstein-Hawking entropy of the resulting four-dimensional black hole, and this holographic agreement can be extended to include arbitrary electric charges as explained in \cite{Hristov:2014hza}. 

Due to the modular properties of two-dimensional conformal field theories it was later shown \cite{Hosseini:2016cyf,Hosseini:2018qsx} that in the Cardy limit the twisted index of ${\cal N}=4$ SYM theory on T$^2 \times$S$^2$ can be evaluated exactly in the gauge group rank $N$, up to exponentially suppressed corrections in one of the fugacities, $\Delta_4$. This therefore gives us the most precise holographic test for the localized QEF so far, where we need to compare it to \eqref{eq:micro_entropy} with
\begin{equation}
 	Z_{{\cal N}=4}^{\mathrm{T}^2\times \mathrm{S}^2} (\mathfrak{p}_a, \Delta_a) = \exp \Bigl[-\frac{16}{27 \Delta_4}\sum_{a=1}^3 \mathfrak{p}_a \frac{\partial\,a_{{\cal N}=4}}{\partial \Delta_a}  + {\cal O} (e^{-1/\Delta_4}) \Bigr]\, ,
\end{equation}
where  $a_{{\cal N}=4}$ is the conformal anomaly
\begin{equation}
		a_{{\cal N}=4} (\Delta_a)  = \frac{27}{32} (N^2-1) \Delta_1 \Delta_2 \Delta_3\, ,
\end{equation}
and the fourth fugacity $\Delta_4$ can be interpreted geometrically as the size of the circle in the dimensional reduction (which only has an electric charge associated with it). Therefore it needs to be taken small, $\Delta_4  \rightarrow 0$, to recover the black hole limit we are interested in. The AdS/CFT dictionary for this model tells us
\begin{equation}
\frac{1}{2\,g^2\,\mathcal{G}_{\rm N}} = \frac{(N^2-1)}{2} \, ,
\end{equation}
up to possible corrections in $1/N$. Using this relation along with \eqref{eq:prepot-BB} and \eqref{eq:FI-BB} for the localized QEF \eqref{eq:dmacro-FT} in this model, one can see the correspondence between macroscopics and microscopics upon relating the fugacities $\Delta_a$ and the integration variables $\phi^I_+$, and the charges as in \eqref{eq:charge-match}. As in the previous example, the saddle-point approximation leads to an exact holographic match. 

\subsection{Gutowski-Reall black holes}
\label{subsec:GRexample}

The Gutowski-Reall black holes \cite{Gutowski:2004ez} and their generalizations \cite{Gutowski:2004yv,Chong:2005da,Chong:2005hr,Kunduri:2006ek} are electrically charged rotating BPS black holes with AdS$_5$ asymptotics. Similarly to the case of black strings, these black holes can also be Scherk-Schwarz reduced to four dimensions \cite{Hosseini:2017mds}, where they fall in the class of static solutions with near-horizon geometry AdS$_2\times$S$^2$ considered in this paper. The resulting four-dimensional solution only has a single fixed magnetic charge, $p^0 = 1$, which also fixes the corresponding dimensionless FI parameter $\widetilde{\xi}_0$. Thus, the supergravity model is specified by
\begin{equation}\label{eq:prepot-GR}
	\mathcal{F}^{\pm}(X_\pm)_{\rm GR} = -\frac{ X^1_{\pm} X^2_{\pm} X^3_{\pm}}{X^0_{\pm}}\, ,
\end{equation}
and dimensionless FI parameters given by
\begin{equation}\label{eq:FI-GR}
	\widetilde{\xi}_0 = \widetilde{\xi}_1 = \widetilde{\xi}_2 = \widetilde{\xi}_3 = \frac12\, .
\end{equation}

Once again we have a type IIB string theory interpretation since these black holes represent supersymmetric states with AdS$_5\times$S$^5$ asymptotics, holographically corresponding to BPS states in the dual ${\cal N}=4$ SYM theory. It was further shown in \cite{Hosseini:2017mds} that the leading macroscopic entropy is computed holographically by the Casimir energy of the SYM theory on a squashed three-sphere \cite{Bobev:2015kza}, which is the large $N$ result for the partition function (superconformal index) of the SYM theory on S$^1 \times$S$^3_{\rm sq}$,
\begin{equation}
\label{eq:Z-GR}
 	Z_{{\cal N}=4}^{\mathrm{S}^1\times \mathrm{S}^3_{\rm sq}} (\mathfrak{p}_0=1, \Delta_a) = \exp\bigl[-E_{{\cal N}=4}(\Delta_a)  + {\cal O} (N^0) \bigr] \, ,
\end{equation}
with the Casimir energy
\begin{equation}
		E_{{\cal N}=4} (\Delta_a)  = \frac{(N^2-1)}{2} \frac{\Delta_1 \Delta_2 \Delta_3}{\Delta_0^2}\, .
\end{equation}
Once again, the holographic relation
\begin{equation}
\frac{1}{2\,g^2\,\mathcal{G}_{\rm N}} = \frac{(N^2-1)}{2} \, ,
\end{equation}
along with \eqref{eq:prepot-GR} and \eqref{eq:FI-GR} establishes an exact correspondence between the leading gravity contribution in \eqref{eq:dmacro-FT} and the field theory expression \eqref{eq:Z-GR}. The subleading corrections need to be computed by evaluating the full superconformal index whose leading behavior in the large $N$ limit is ${\cal O} (N^0)$. Therefore at the moment we can only correctly reproduce the saddle-point match between the macroscopic and microscopic entropy and look forward to the development of new tools for the exact evaluation of the superconformal index (see \cite{Dolan:2011rp,Spiridonov:2012ww} and references therein for work in this direction).

\subsection{Massive type IIA black holes}
\label{subsec:Massiveexample}

Another genuinely AdS$_4$ holographic example was more recently analyzed in \cite{Guarino:2017eag,Guarino:2017pkw,Hosseini:2017fjo,Benini:2017oxt}, where black holes in maximal 4d $ISO(7)$-gauged supergravity were considered. Via the uplift of \cite{Guarino:2015vca} these solutions can be embedded in massive type IIA theory and holographically correspond to D2$_k$ branes \cite{Guarino:2015jca} (3d $\mathrm{SU}(N)$ SYM theory with Chern-Simons interaction at level $k$ corresponding to the Romans' mass) wrapped on a two-cycle. The consistent truncation to a 4d ${\cal N}=2$ gauged supergravity model \cite{Guarino:2017pkw} includes a hypermultiplet gauging, but via the supersymmetry-preserving Higgs mechanism it was shown \cite{Hosseini:2017fjo} that the model can be further truncated to the FI gauged supergravity considered here. The resulting black hole solutions have three electromagnetic charges and the model is given by the prepotential
 \begin{equation}\label{eq:prepot-IIA}
	\mathcal{F}^{\pm}(X_\pm)_{\mathrm{mIIA}} = -\frac{3^{3/2}}{4} (-c)^{1/3} (X_\pm^1 X_\pm^2 X_\pm^3)^{2/3}\, ,
\end{equation}
where the parameter $c$ is related to the Romans' mass $k$. The dimensionless FI parameters are given by
\begin{equation}\label{eq:FI-IIA}
	\widetilde{\xi}_1 = \widetilde{\xi}_2 = \widetilde{\xi}_3 = 1\, .
\end{equation}

The holographic calculation proceeds via the large $N$ limit of the twisted index of the D2$_k$ theory, computed using localization. Unfortunately at the moment there are no known results for the subleading corrections to the partition function in this setting, and therefore only the saddle-point match has been successfully performed after the coupling constants in the dual theories are carefully mapped to each other. The large $N$ answer for the twisted index gives
 \begin{equation}
 	Z_{D2_k} (\mathfrak{p}_a, \Delta_a) = \exp \Bigl[-\sum_{a=1}^3 \mathfrak{p}_a \frac{\partial {\cal F}_{D2_k}}{\partial \Delta_a}  + {\cal O} (N \log N) + {\cal O} (N^{2/3})\Bigr]\, ,
\end{equation}
with ${\cal F}_{D2_k}$ is the partition function of the theory on S$^3$ \cite{Fluder:2015eoa},
\begin{equation}
		{\cal F}_{D2_k} (\Delta_a) = -\frac{3^{13/6}}{5\times 2^{5/3}} (-k)^{1/3} N^{5/3} (\Delta_1 \Delta_2 \Delta_3)^{2/3}\, .
\end{equation}
The holographic relations
\begin{equation}
\frac{c^{1/3}}{2\,g^2\,\mathcal{G}_{\rm N}} = \frac{2^{1/3}\ 3^{2/3}}{5} \, k^{1/3}\, N^{5/3}\, , \qquad c = \left(\frac{3}{16 \pi^3}\right)^{1/5} k\, N^{1/5}\, , 
\end{equation}
together with \eqref{eq:prepot-IIA} and \eqref{eq:FI-IIA} again shows an exact match in the leading gravity and field theory answers.
Due to the similarities in the evaluation of the partition function here and in the example discussed in subsection \ref{subsec:CKexample}, it is natural to expect that also here the $\log N$ contribution has a simple form independent of the chemical potentials $\Delta_a$. We are however not aware of any explicit attempts to tackle finite $N$ corrections in the D2$_k$ theory.   

\section{Conclusion and discussion}
\label{sec:conclusion}

In this paper, we have completed an important first step in the evaluation of the exact degeneracy of horizon degrees of freedom for a class of half-BPS black hole attractors with AdS asymptotics. Localization techniques yield the result \eqref{eq:dmacro-loc-reg} (or \eqref{eq:dmacro-FT} in variables which are suited for holography) for the quantum entropy function \eqref{eq:macroscopic-degeneracy}. Using this result, we have been able to exhibit a perfect match in the large $N$ limit with the results obtained in the dual field theories for a number of examples.

Such a check is of course necessary, but perhaps more importantly the results derived here also provide a clear path for the analysis of finite $N$ corrections. In particular, according to \eqref{eq:dmacro-FT}, all such corrections (modulo the presence of hair degrees of freedom) are captured in the gravitational theory by the quantity $Z_{\rm ind}^{\rm reg}$, which combines the effects of the usual one-loop determinants in localization together with bosonic and fermionic zero-modes on the localization locus, and an eventual non-trivial measure along this locus. It is encouraging that a number of techniques have already been put forward in other situations (and in particular for asymptotically flat black holes) which should allow for a straightforward, albeit technically challenging, computation of this quantity. This is clearly the main future research interest, and we hope that progress in this direction will pave the way for a prediction of finite $N$ effects directly from the gravitational side of the duality. Ultimately, this could lead to finite $N$ precision holography and deeper insights into the structure of quantum gravity. 

Aside from these considerations, there are a few other aspects of this work that deserve further attention and will also be of future research interest. As already stated throughout this paper, the main assumption we started with is the absence of off-shell fluctuations for the Weyl multiplet. This assumption effectively ``freezes'' the metric and related supergravity fields to their on-shell values. This seems to be a valid assumption \emph{a posteriori}, as it leads to an agreement between the localizing configurations for the vector multiplet and hypermultiplet sectors. It is tempting to speculate that our final result is insensitive to potential off-shell fluctuations in the Weyl multiplet sector, but clearly this question deserves careful further investigation.

Interestingly, we were able to evaluate the classical action of supergravity on the localization locus regardless of the explicit form of the prepotential. The locus a priori depends on $2 n_V + 2$ sets of infinitely many free fluctuation functional parameters $\{C^I_k(\theta,\varphi), D^I_k(\theta,\varphi)\}$ for all $k\geq 1$. The final integrand, however, only depends on $n_V$ independent linear combinations of these variables, and the BPS equations further imply that precisely such linear combinations are constant on the 2-sphere. This essentially means that we could have derived the final result \eqref{eq:dmacro-loc-reg} only assuming that a single entry in the set $\{C^I_k(\theta,\varphi), D^I_k(\theta,\varphi)\}$ is nonzero and independent of the angular coordinates for a given $I$. This also relates to our assumption that the directions orthogonal to the $n_V$ linear combinations on the localization locus yield a finite factor when properly regularized by taking fermionic zero-modes into account. In contrast with the analysis of asymptotically flat black holes, the integral over these orthogonal directions is still a functional integral due to the angular dependence. This feature is specific to the half-BPS attractor background we considered in this paper. We expect that a detailed analysis of the factor $Z_{\rm ind}^{\rm reg}$ will offer insights into the resolution of these issues.

\section*{Acknowledgements}
We would like to thank Bernard de Wit, Seyed Morteza Hosseini, Noppadol Mekareeya, Sameer Murthy, Paul Richmond and Alberto Zaffaroni for numerous discussions. KH is supported in part by the Bulgarian NSF grant DN08/3 and the bilateral grant STC/Bulgaria-France 01/6. VR is supported in part by INFN and by the ERC Starting Grant 637844-HBQFTNCER. VR would also like to acknowledge the hospitality of the Bulgarian Academy of Sciences, Sofia, where part of this work was conducted.

\appendix

\section{Supergravity details}
\label{app:sugra-details}

\subsection{Conventions}
\label{app:conventions}

Space-time indices are denoted by Greek letters $\mu,\nu,\ldots$ while tangent space indices are denoted by Roman letters $a,b,\ldots$. (Anti-)symmetrization of indices is always done with total weight one, e.g.
\begin{equation}
X_{[a}X_{b]} := \tfrac12\,(X_a X_b - X_b X_a) \, .
\end{equation}

The dual of a rank-2 tensor in four Euclidean dimensions is defined as
\begin{equation}
\widetilde{T}_{ab} = \tfrac12\,\varepsilon_{abcd}\,T^{cd} \, , \qquad \mathrm{with} \quad \varepsilon_{1234} = \varepsilon^{1234} = 1 \, .
\end{equation}
The (anti-)self-dual part of such a tensor is defined as
\begin{equation}
T_{ab}^\pm = \tfrac12(T_{ab} \pm \widetilde{T}_{ab}) \, . \vspace{2mm}
\end{equation}

In four Euclidean dimensions, the Clifford algebra is
\begin{equation}
\{\gamma^a,\,\gamma^b\} = 2\,\delta^{ab} \, .
\end{equation}
We use Hermitian $\gamma$-matrices, and define
\begin{equation}
\gamma^5 := \gamma_1\,\gamma_2\,\gamma_3\,\gamma_4 \, .
\end{equation}
When an explicit Hermitian representation of the Clifford algebra is needed, we will use
\begin{equation}
\label{eq:gamma-basis}
\gamma_1 = \sigma_1 \otimes \oneone \, , \quad \gamma_2 = \sigma_2 \otimes \oneone \, , \quad \gamma_3 = \sigma_3 \otimes \sigma_1 \, , \quad \gamma_4 = \sigma_3 \otimes \sigma_2 \, ,
\end{equation}
where $\sigma_i$, $i=1,2,3$ are the Pauli matrices.\\

We define the Dirac conjugate of a spinor as
\begin{equation}
\bar{\psi}_i := (\psi^i)^\dagger \, ,
\end{equation}
where complex conjugation raises and lowers $\mathrm{SU}(2)_\mathrm{R}$ indices. The symplectic-Majorana reality condition reads
\begin{equation}
  \label{eq:symp-Majo}
C^{-1}\,\bar{\psi}_i{}^\mathrm{T} = \varepsilon_{ij}\,\psi^j \, , 
\end{equation}
where the charge conjugation matrix is anti-symmetric, unitary, and such that\footnote{In the explicit basis \eqref{eq:gamma-basis}, we take $C\gamma^5 = \sigma_1 \otimes \sigma_2$.}
\begin{equation}
C\,\gamma_a\,C^{-1} = - \gamma_a{}^\mathrm{T} \, , \quad C\,\gamma^5\,C^{-1} = \gamma^5 \, ,
\end{equation}
while the anti-symmetric tensor of $\mathrm{SU}(2)$ satisfies
\begin{equation}
\varepsilon^{ij}\,\varepsilon_{jk} = -\delta^i{}_k \, , \quad \mathrm{and} \quad \varepsilon^{ij}\,\varepsilon_{ij} = 2 \, .
\end{equation}
We note that \eqref{eq:symp-Majo} also applies to the chiral projections
\begin{align}
\psi^i_\pm :=& \tfrac12(\mathbb{1} \pm \gamma^5)\,\psi^i \, . \\ \nonumber
\end{align}

For fermionic bilinears with \emph{anti-commuting} spinors fields and a matrix $\Gamma$ built out of products of $\gamma$-matrices, we note the following result,
\begin{equation}
\label{eq:Majo-flip}
(\bar{\phi_j}\,\Gamma^\dagger\,\psi^i)^\dagger = \bar{\psi}_i\,\Gamma\,\phi^j = -\varepsilon_{ik}\,\varepsilon^{jl}\,\bar{\phi}_l\,C^{-1}\,\Gamma^\mathrm{T}\,C\,\psi^k \, .
\end{equation}
For \emph{commuting} spinor fields, the last equality acquires an extra minus sign. For completeness, we also give the Fierz rearrangement formula for anti-commuting chiral spinors,
\begin{align}
\phi^j_\pm\,\bar{\psi}_{i\pm} =&\; -\tfrac14(\mathbb{1} \pm \gamma^5)\,(\bar{\psi}_{i\pm}\,\phi^j_\pm) + \tfrac18\,\gamma^{ab}\,(\bar{\psi}_{i\pm}\,\gamma_{ab}\,\phi^j_\pm) \, , \\
\phi^j_\mp\,\bar{\psi}_{i\pm} =&\; - \tfrac14\,\gamma^a\,(\mathbb{1} \pm \gamma^5)\,(\bar{\psi}_{i\pm}\,\gamma_a\,\phi^j_\mp) \, .
\end{align}

\subsection{Locally supersymmetric Euclidean Lagrangian densities}
\label{app:actions}

Using the Euclidean multiplet calculus \cite{deWit:2017cle}, one can build a superconformally invariant Lagrangian density for an arbitrary number of vector multiplets coupled to the Weyl multiplet. This density is based on two prepotentials $\mathcal{F}^\pm(X_\pm)$, which are homogeneous functions of degree two in the scalar fields of the vector multiplets,
\begin{equation}
\mathcal{F}^\pm(\lambda^2 X_\pm^I) = \lambda^2\,\mathcal{F}^\pm(X_\pm^I) \, .
\end{equation}
We note that this homogeneity also implies
\begin{equation}
X^I_\pm\,\mathcal{F}_I^\pm = 2\,\mathcal{F}^\pm \, , \quad \mathrm{and} \quad X^J_\pm\,\mathcal{F}_{IJ}^\pm = \mathcal{F}_I^\pm \, ,
\end{equation}
where the indices on the prepotentials denote derivatives with respect to the argument,
\begin{equation}
\mathcal{F}_I^\pm := \frac{\partial\,\mathcal{F}^\pm(X_\pm)}{\partial\,X^I_\pm} \, , \quad \mathrm{and} \quad \mathcal{F}_{IJ}^\pm := \frac{\partial^2\,\mathcal{F}^\pm(X_\pm)}{\partial\,X^I_\pm\,\partial\,X^J_\pm} \, .
\end{equation}
The bosonic terms of the vector multiplet Lagrangian density are given by
\begin{align}
  \label{eq:vector-action}
e^{-1}&\mathcal{L}_\mathrm{V} = -\bigl(X^I_+\,\mathcal{F}^-_I + X^I_-\,\mathcal{F}^+_I\bigr)\bigl(\tfrac16\,R - D\bigr) + \bigl(\mathcal{F}_{IJ}^+ + \mathcal{F}_{IJ}^-\bigr)\mathcal{D}_\mu X^I_+\mathcal{D}^\mu X^J_- \nonumber \\[1mm]
&+ \tfrac14\,\mathcal{F}_{IJ}^+\bigl[F_{ab}^{-\,I} - \tfrac14 X^I_- T_{ab}^-\bigr]\bigl[F^{ab-J} - \tfrac14 X^J_- T^{ab-}\bigr] + \tfrac14\,\mathcal{F}_{IJ}^-\bigl[F_{ab}^{+\,I} - \tfrac14 X^I_+T_{ab}^+\bigr]\bigl[F^{ab+J} - \tfrac14 X^J_+ T^{ab+}\bigr] \nonumber \\[1mm]
&- \tfrac18\,\mathcal{F}^+_I\bigl[F_{ab}^{+\,I} - \tfrac14\,X^I_+T_{ab}^+\bigr]T^{ab+} - \tfrac18\,\mathcal{F}^-_I\bigl[F_{ab}^{-\,I} - \tfrac14\,X^I_- T_{ab}^-\bigr]T^{ab-} \nonumber \\[1mm]
&+ \tfrac18\,\bigl(\mathcal{F}_{IJ}^+ + \mathcal{F}_{IJ}^-\bigr)\,Y_{ij}^I\;Y^{ij\,J} - \tfrac1{32}\,\mathcal{F}^+(T_{ab}^+)^2 - \tfrac{1}{32}\,\mathcal{F}^-(T_{ab}^-)^2 \, .
\end{align}
In the main body of the paper, we will often use the standard notations
\begin{equation}
\label{eq:Kahler-pot}
\chi_\mathrm{V} := X^I_+\,\mathcal{F}^-_I + X^I_-\,\mathcal{F}^+_I \, , \quad \mathrm{and} \quad N_{IJ} := \mathcal{F}_{IJ}^+ + \mathcal{F}_{IJ}^- = \frac{\partial^2\,\chi_\mathrm{V}}{\partial X^I_+\,\partial X^J_-} \, .
\end{equation}

The locally supersymmetric Euclidean Lagrangian density for hypermultiplets transforming under a certain local gauge group contains the following bosonic terms,
\begin{align}
  \label{eq:hyper-action}
e^{-1}\mathcal{L}_\mathrm{H} =&\; \tfrac12\,\varepsilon^{ij}\,\Omega_{\alpha\beta}\,A_i{}^\alpha A_j{}^\beta\bigl(\tfrac16\,R + \tfrac12\,D\bigr) - \tfrac12\,\varepsilon^{ij}\,\Omega_{\alpha\beta}\,\mathcal{D}_\mu A_i{}^\alpha\,\mathcal{D}^\mu A_j{}^\beta \nonumber \\[1mm]
&\;+ 2\, \Omega_{\alpha\beta}\varepsilon^{ij} A_i{}^\alpha\,\xi_I X^I_-\,t^\beta{}_\gamma\,\xi_J X^J_+\,t^\gamma{}_\delta\,A_j{}^\delta - \tfrac12\, \Omega_{\alpha\beta}A_i{}^\alpha\,\xi_I Y^{ij\,I}\,t^\beta{}_\gamma\,A_j{}^\gamma \, ,
\end{align}
with $\xi_I$ the FI parameters and $t^\alpha{}_\beta$ the generators of the gauging. We will also write
\begin{equation}
\chi_\mathrm{H} := \tfrac12\,\varepsilon^{ij}\,\Omega_{\alpha\beta}\,A_i{}^\alpha\,A_j{}^\beta \, .
\end{equation}

\section{Attractor configuration details}
\label{app:attract-details}

\subsection{Conformal Killing spinors of the attractor geometry}
\label{app:CKS}

In this Appendix, we derive the explicit conformal Killing spinors parameterizing the unbroken supercharges of the bosonic near-horizon attractor configuration presented in sections \ref{subsec:Weyl-attractor} and \ref{subsec:Matter-attractor}. All the equations we write here pertain to the background, and we will therefore drop the circle on the matter fields for ease of presentation.

Following \cite{deWit:2011gk}, the strategy to find the conformal Killing spinors is to consider an $S$-supersymmetry invariant combination of spinors built out of the gravitino and gaugino fields. To do so, we introduce
\begin{equation}
\Omega_\pm^{i\,\mathrm{V}} := \tfrac12\,\chi_\mathrm{V}^{-1}\,X^I_\mp\,N_{IJ}\,\Omega_\pm^{i\,J} \, ,
\end{equation}
where the $\pm$ subscript on fermions refers to chirality and $\chi_\mathrm{V}$ is the K\"{a}hler potential \eqref{eq:Kahler-pot}. In the bosonic background, we have
\begin{equation}
\delta\Omega_\pm^{i\,\mathrm{V}} = \mathcal{A}^i{}_{j\,\mp}\,\epsilon^j_\pm + \eta^i_\pm \, ,
\end{equation}
where
\begin{equation}
\mathcal{A}^i{}_{j\,\mp} := -\tfrac12\,\chi_{\mathrm{V}}^{-1}\,N_{IJ}\,X^I_\mp\,\varepsilon_{kj}\,Y^{ik\,J} \, ,
\end{equation}
Using the equations of motion for $Y^I_{ij}$ and the auxiliary scalar $D$, we can write
\begin{equation}
\mathcal{A}^i{}_{j\,\mp} = 2\,X^I_\mp\,\mu_I{}^{ik}\,\varepsilon_{kj} \, ,
\end{equation}
where $\mu_{ij\,I}$ are the moment maps \eqref{eq:1/2-BPS-moment-maps}. We now consider the variation of the $S$-invariant combination
\begin{equation}
\delta\bigl(\psi_{\mu\,\pm}^i + \mathrm{i}\,\gamma_\mu\Omega_\mp^{i\,\mathrm{V}}\bigr) = 2\,\nabla_\mu\epsilon^i_\pm + \mathcal{V}_\mu{}^i{}_j\,\epsilon^j_\pm \mp \mathrm{i}\,\Bigl(\frac{1}{2\sqrt{v_1}}\,\gamma^{34}\gamma^5\delta^i{}_j \mp \mathcal{A}^i{}_{j\,\pm}\Bigr)\gamma_\mu\epsilon^j_\mp \, ,
\end{equation}
where $\nabla_\mu$ denotes the covariant derivative on ${\cal H}_2\times {\rm S}^2$, which contains only the spin-connection in the gauge-fixed background \eqref{eq:K-A-gauge-fix}. Setting the above variation to zero yields the (generalized) Killing spinor equation
\begin{equation}
\label{eq:1/2-BPS-KS}
2\,\nabla_\mu\,\epsilon^i_\pm + \mathcal{V}_\mu{}^i{}_j\,\epsilon^j_\pm + \frac{1}{2\sqrt{v_1}}\Bigl(\sigma_3{}^i{}_j - \mathrm{i}\,\gamma^{34}\delta^i{}_j\Bigr)\gamma_\mu\,\epsilon^j_\mp = 0 \, .
\end{equation}
To solve this equation, let us examine it on the two factors of the ${\cal H}_2 \times {\rm S}^2$ geometry, where we split the space-time index $\mu = (\underline{\mu},\,\hat{\mu})$ accordingly.\\

The $\mathrm{SU}(2)_\mathrm{R}$ gauge field only has components along the ${\rm S}^2$ according to \eqref{eq:1/2-BPS-SU(2)-connection}. Thus, on the ${\cal H}_2$ factor, \eqref{eq:1/2-BPS-KS} reduces to\footnote{Here we only display the $i\!=\!1$ component of the symplectic Majorana doublet, since the $i\!=\!2$ component is obtained by imposing the reality condition \eqref{eq:symp-Majo}.}
\begin{equation}
2\,\nabla_{\underline{\mu}}\,\epsilon^1_\pm + \frac{1}{\sqrt{v_1}}\,\gamma_{\underline{\mu}}\,\xi^1_\mp = 0 \, ,
\end{equation}
where we introduced the projected spinor
\begin{equation}
\xi^1 := \frac12\bigl(\oneone - \mathrm{i}\,\gamma^{34})\,\epsilon^1 \, .
\end{equation}

We now impose the following half-BPS projection on the spinor parameter \cite{deWit:2011gk}: 
\begin{equation}
\label{eq:1/2-BPS-proj}
\gamma^{34}\epsilon^i_\pm = \mathrm{i}\,\sigma_3{}^i{}_j\,\epsilon^j_\pm \, .
\end{equation}
As a consequence, the unbroken supersymmetries of the attractor background are parameterized by $\xi^1_+$ and $\xi^1_-$, which satisfy the equation
\begin{equation}
\label{eq:KS-AdS}
2\,\nabla_{\underline{\mu}}\,\xi^1_\pm + \frac{1}{\sqrt{v_1}}\,\gamma_{\underline{\mu}}\,\xi^1_\mp = 0 \, .
\end{equation}

A similar analysis on the ${\rm S}^2$ factor shows that the parameters of the unbroken supersymmetries satisfy
\begin{equation}
2\,\nabla_{\hat{\mu}}\,\xi^1_\pm + \mathcal{V}_{\hat{\mu}}{}^1{}_1\,\xi^1_\pm = 0 \, .
\end{equation}
These equations are not the standard equations for Killing spinors on the 2-sphere \cite{deWit:2011gk}, but we can use the $\mathrm{SU}(2)_\mathrm{R}$ connection to effect a twist and cancel the spin-connection present in the covariant derivative. For $\hat{\mu} = \theta$, both the spin-connection and the R-connection vanish on the attractor background, and we are left simply with
\begin{equation}
\partial_\theta\,\xi^1_+ = \partial_\theta\,\xi^1_- = 0 \, .
\end{equation}
For $\hat{\mu} = \varphi$, we obtain instead
\begin{equation}
2\,\partial_\varphi\,\xi^1_\pm - \cos\theta\,\gamma^{34}\,\xi^1_\pm + 2\mathrm{i}\,\xi_I p^I\cos\theta\,\xi^1_\pm = 0 \, .
\end{equation}
With our choice of gauging \eqref{eq:elec-gauging}, this is
\begin{equation}
2\,\partial_\varphi\,\xi^1_\pm + \mathrm{i}\cos\theta\,\bigl(\oneone + \mathrm{i}\,\gamma^{34}\bigr)\,\xi^1_\pm = 0 \, .
\end{equation}
The second term vanishes owing to the half-BPS projection \eqref{eq:1/2-BPS-proj}, and we are left with
\begin{equation}
\partial_\varphi\,\xi^1_+ = \partial_\varphi\,\xi^1_- = 0 \, .
\end{equation}
In conclusion, the Killing spinors for the unbroken supersymmetries of the bosonic on-shell half-BPS attractor background are simply constant on the ${\rm S}^2$. On ${\cal H}_2$, they satisfy the generalized Killing spinor equation \eqref{eq:KS-AdS}.\\

To solve explicitly for $\xi^1_\pm$, we use an Ansatz compatible with chirality and the half-BPS projection. Using an explicit $\gamma$-matrix basis (see \eqref{eq:gamma-basis}), we write
\begin{equation}
\label{eq:xi-1-ansatz}
\xi^1_+ =\begin{pmatrix} 0 \\ g_+(\tau,\eta) \\ 0 \\ 0 \end{pmatrix} \, , \quad \xi^1_- = \begin{pmatrix} g_-(\tau,\eta) \\ 0 \\ 0 \\ 0 \end{pmatrix} \, .
\end{equation}
The (generalized) Killing spinor equation \label{eq:1/2-BPS-KS-AdS} then implies the following first-order coupled partial differential equations on $g_+$ and $g_-$
\begin{align}
0 =&\; 2\,\partial_\tau g_+ - \mathrm{i}\cosh\eta\,g_+ + \sinh\eta\,g_- \, , \\
0 =&\; 2\,\partial_\tau g_- + \mathrm{i}\cosh\eta\,g_- + \sinh\eta\,g_+ \, , \\
0 =&\; 2\,\partial_\eta g_+ + \mathrm{i}\,g_- \, , \\
0 =&\; 2\,\partial_\eta g_- - \mathrm{i}\, g_+ \, .
\end{align}
Solving for $g_+$ and $g_-$, we obtain the following expression for $\xi^1_\pm$,
\begin{equation}
\label{eq:xi-1}
\xi^1_+ =\begin{pmatrix} 0 \\ \alpha\,e^{\frac12\mathrm{i}\,\tau}\cosh\bigl(\frac{\eta}{2}\bigr) - \mathrm{i}\,\beta\,e^{-\frac12\mathrm{i}\,\tau}\sinh\bigl(\frac{\eta}{2}\bigr) \\ 0 \\ 0 \end{pmatrix} , \; \xi^1_- = \begin{pmatrix} \beta\,e^{-\frac12\mathrm{i}\,\tau}\cosh\bigl(\frac{\eta}{2}\bigr) + \mathrm{i}\,\alpha\,e^{\frac12\mathrm{i}\,\tau}\sinh\bigl(\frac{\eta}{2}\bigr) \\ 0 \\ 0 \\ 0 \end{pmatrix} ,
\end{equation}
where $\alpha$ and $\beta$ are arbitrary complex constants.

Taking the $\gamma$-trace of the original gravitino variation \eqref{eq:gravitino-var} now gives the parameters of the unbroken $S$-supersymmetries on the background,
\begin{equation}
\label{eq:eta-epsilon}
\kappa^1_\pm := \frac12\bigl(\oneone - \mathrm{i}\,\gamma^{34})\,\eta^1_\pm = -\frac12\mathrm{i}\,\Slash{\mathcal{D}}\,\xi^1_\mp \, .
\end{equation}
Explicitly,
\begin{equation}
\label{eq:kappa-1}
\kappa^1_+ = \begin{pmatrix} 0 \\ \frac{\mathrm{i}\,\alpha\,e^{\frac12\mathrm{i}\,\tau}\cosh\bigl(\tfrac{\eta}{2}\bigr) + \beta\,e^{-\frac12\mathrm{i}\,\tau}\sinh\bigl(\tfrac{\eta}{2}\bigr)}{2\sqrt{v_1}} \\ 0 \\ 0 \end{pmatrix} , \; \kappa^1_- = \begin{pmatrix} \frac{\mathrm{i}\,\beta\,e^{-\frac12\mathrm{i}\,\tau}\cosh\bigl(\tfrac{\eta}{2}\bigr) - \alpha\,e^{\frac12\mathrm{i}\,\tau}\sinh\bigl(\tfrac{\eta}{2}\bigr)}{2\sqrt{v_1}} \\ 0 \\ 0 \\ 0 \end{pmatrix} .
\end{equation}
Lastly, we can find the $i=2$ components of the symplectic Majorana doublets by imposing the symplectic Majorana condition \eqref{eq:symp-Majo}:
\begin{equation}
\label{eq:xi-2}
\xi^2_+ =\begin{pmatrix} 0 \\ 0 \\ \mathrm{i}\bar{\alpha}\,e^{-\frac12\mathrm{i}\,\tau}\cosh\bigl(\frac{\eta}{2}\bigr) - \bar{\beta}\,e^{\frac12\mathrm{i}\,\tau}\sinh\bigl(\frac{\eta}{2}\bigr) \\ 0 \end{pmatrix} , \; \xi^2_- = \begin{pmatrix} 0 \\ 0 \\ 0 \\ -\mathrm{i}\,\bar{\beta}\,e^{\frac12\mathrm{i}\,\tau}\cosh\bigl(\frac{\eta}{2}\bigr) - \bar{\alpha}\,e^{-\frac12\mathrm{i}\,\tau}\sinh\bigl(\frac{\eta}{2}\bigr) \vspace{2mm} \end{pmatrix} ,
\end{equation}
and
\begin{equation}
\label{eq:kappa-2}
\kappa^2_+ = \begin{pmatrix} 0 \\ 0 \\ \frac{\bar{\alpha}\,e^{-\frac12\mathrm{i}\,\tau}\cosh\bigl(\tfrac{\eta}{2}\bigr) + \mathrm{i}\,\bar{\beta}\,e^{\frac12\mathrm{i}\,\tau}\sinh\bigl(\tfrac{\eta}{2}\bigr)}{2\sqrt{v_1}} \\ 0 \end{pmatrix} , \; \kappa^2_- = \begin{pmatrix} -\frac{\bar{\beta}\,e^{\frac12\mathrm{i}\,\tau}\cosh\bigl(\tfrac{\eta}{2}\bigr) - \mathrm{i}\,\bar{\alpha}\,e^{-\frac12\mathrm{i}\,\tau}\sinh\bigl(\tfrac{\eta}{2}\bigr)}{2\sqrt{v_1}} \\ 0 \\ 0 \\ 0 \end{pmatrix} .
\end{equation}

In conclusion, the on-shell attractor background preserves four conformal supercharges, encoded in the two complex chiral charges
\begin{equation}
\mathcal{Q}_+ := (\xi^i_+)^\dagger Q^i + (\kappa^i_+)^\dagger S^i \, , \quad \mathrm{and} \quad \mathcal{Q}_- := (\xi^i_-)^\dagger Q^i + (\kappa^i_-)^\dagger S^i \, .
\end{equation}
In the main text, we will make use of the combinations
\begin{equation}
\mathcal{Q} := \mathcal{Q}_+ + \mathcal{Q}_- \, , \quad \mathrm{and} \quad \widetilde{\mathcal{Q}} := \mathcal{Q}_+ - \mathcal{Q}_- \, ,
\end{equation}
which we parameterize using \emph{commuting} spinors $\xi^i := \xi^i_+ + \xi^i_-$ and $\widetilde{\xi}^{\,i} := \xi^i_+ - \xi^i_-$ and the corresponding $\kappa^i$ and $\widetilde{\kappa}^{\,i}$.

\subsection{Isometry superalgebra of the attractor geometry}
\label{app:superalgebra}

The superconformal symmetry of the on-shell half-BPS attractor background presented in section \ref{sec:on-shell} is the direct product $\mathfrak{su}(1,1|1) \times \mathfrak{su}(2)$. It contains the bosonic subalgebra
\begin{equation}
\mathfrak{su}(1,1) \times \mathfrak{u}(1)_\mathrm{R} \times \mathfrak{su}(2) \, ,
\end{equation}
where the first factor is identified with the isometries of ${\cal H}_2$, the second with the R-symmetry group, and the third with the S$^2$ isometries. We denote the generators of $\mathfrak{su}(1,1)$ by $\{L_0,L_{\pm 1}\}$, and the generator of the $\mathfrak{u}(1)_\mathrm{R}$ factor by $R$. The odd elements of the superalgebra are the superconformal symmetries $G_{\pm \frac12}$ and $\overline{G}_{\pm \frac12}$, in accordance with standard notation, cf.\ e.g.\ App.\ B in \cite{Benini:2015eyy}. Their non-vanishing (anti-)commutation relations are
\begin{align}
\bigl[L_0,\,L_{\pm 1}\bigr] =&\; \mp L_{\pm 1} \, , \qquad \;\; \quad \bigl[L_{+1},\,L_{-1}\bigr] = 2\,L_0 \, , \\
\bigl[L_0,\,G_{\pm \frac12}\bigr] =&\; \mp\tfrac12\,G_{\pm \frac12} \, , \qquad \; \bigl[L_{\pm 1},\,G_{\mp \frac12}\bigr] = \pm\,G_{\pm \frac12} \, , \\[1mm]
\bigl[L_0,\,\overline{G}_{\pm \frac12}\bigr] =&\; \mp\tfrac12\,\overline{G}_{\pm \frac12} \, , \qquad \; \bigl[L_{\pm 1},\,\overline{G}_{\mp \frac12}\bigr] = \pm\,\overline{G}_{\pm \frac12} \, , \\[1mm]
\bigl[R,\,G_{\pm \frac12}\bigr] =&\; G_{\pm \frac12} \, , \qquad \qquad \; \bigl[R,\,\overline{G}_{\pm \frac12}\bigr] = -\,\overline{G}_{\pm \frac12} \, , \\[1mm]
\bigl\{G_{\pm \frac12},\,\overline{G}_{\pm \frac12}\bigr\} =&\; 2\,L_{\pm 1} \, , \qquad \qquad \bigl\{G_{\pm \frac12},\,\overline{G}_{\mp \frac12}\bigr\} = 2\, L_0 \pm R \, .
\end{align}
The hermiticity properties are $L_0^\dagger = L_0, \, L_{\pm 1}^\dagger = L_{\mp 1}, \, R^\dagger = R, $ and $G_{\pm \frac12}^\dagger = \overline{G}_{\mp \frac12}$.

From this algebra, we see that the supercharge ${\cal G}_+ := G_{\frac12} + \overline{G}_{-\frac12}$ squares to $2L_0 + R$, while ${\cal G}_- := G_{-\frac12} + \overline{G}_{\frac12}$ squares to $2L_0 - R$, giving in both cases a compact symmetry ($L_0$ is a compact isometry of ${\cal H}_2$ while $R$ is the $U(1)_R$ group). Therefore we can freely choose either of the ${\cal G}_\pm$ as our localizing supercharge. According to the background algebra presented in \eqref{eq:loc-alg-QQ}, ${\cal G}_+$ corresponds to $\mathcal{Q}$ with $\beta=0$ in the $(\xi^i,\kappa^i)$ parameters of App. \ref{app:CKS}, while ${\cal G}_-$ corresponds to $\mathcal{Q}$ with $\alpha=0$ in the $(\xi^i,\kappa^i)$ parameters.

\providecommand{\href}[2]{#2}

\end{document}